\documentclass[12pt,preprint]{aastex}

\shorttitle{On the origin of 3 seismic sources }
\shortauthors{Zharkova \& Zharkov}

\begin{document}

\title{On the origin of 3 seismic sources in the proton-rich flare of October 28, 2003 }

\author{Valentina V. Zharkova\altaffilmark{1} and Sergei I. Zharkov \altaffilmark{2}}

\altaffiltext{1}{Department of Computing, University of Bradford, Bradford, UK; v.v.zharkova@brad.ac.uk}

\altaffiltext{2}{Department of Applied Mathematics, University of Sheffield, Sheffield, UK; s.zharkov@sheffield.ac.uk }

\begin{abstract}

The 3 seismic sources S1, S2 and S3 detected from MDI dopplergrams using the time-distance diagram technique are presented with the locations, areas and vertical and horizontal velocities of the visible wave displacements. Within the datacube of 120 Mm the horizontal velocities and the wave propagation times slightly vary from source to source. The momenta and start times measured from the TD diagrams in the sources S1-S3 are compared with those delivered to the photosphere by different kinds of high energy particles with the parameters deduced from hard X-ray and $\gamma$-ray emission as well as by the hydrodynamic shocks caused by these particles. The energetic protons (power laws combined with quasi-thermal ones, or jets) are shown to deliver momentum high enough and to form the hydrodynamic shocks deeply in a flaring atmosphere that allows them to be delivered to the photosphere through much shorter distances and times. Then the seismic waves observed in the sources S2 and S3 can be explained by the momenta produced by hydrodynamic shocks which are caused by mixed proton beams and jets occurring nearly simultaneously with the third burst of hard X-ray (HXR) and $\gamma$-ray emission in the loops with footpoints in the locations of these sources. The seismic wave in the source S1, delayed by 4 and 2 minutes from the first and second HXR bursts, respectively, is likely to be associated with a hydrodynamic shock occurring in this loop from precipitation of a very powerful and hard electron beam with higher energy cutoff mixed with quasi-thermal protons generated by either of these 2 bursts.

\end{abstract} 


\section{Introduction} \label{intro}
The first successful attempt to observe seismic waves in the SOHO/MDI dopplergrams in a form of ripples  centred on the  X2.6/1B solar flare of 9th July 1996 was reported by Kosovichev $\&$ Zharkova (1998). The authors presented a helioseismic response (solar quake) propagating on the solar surface from the flare location  with a strong localised plasma downflow of about 1.5 km/s in the location of the $H_{\alpha}$ flare impulse occurring within a minute close to the hard X-ray maximum \citep{kozh98}. 

A comparison of the observations with the theoretical model \cite{kozh95} revealed that the momentum required to produce the observed seismic response ($\sim 2\cdot 10^{22}$ $g\, cm\, s^{-1}$) is one order of magnitude higher than those of $\sim 10^{21}$ $g\, cm\, s^{-1}$  observed from the plasma downflows in the MDI dopplergrams \cite{kozh98}. The required momentum could be delivered by a hydrodynamic shock appearing at the injection of a very hard ($\gamma=3$) and intense ($F_0=10^{12}$ $erg/cm^2/s$) electron beam. These parameters were arbitrary selected since there are no detailed hard X-ray observations available for this flare. 

However, the travel time of this shock to the photosphere is more than 2 minutes while the time, at which the helioseismic response started in TD diagrams, coincides very closely with the time of the hard X-ray impulse and does not reveal the $2$ minute delay. Hence, there should be some additional sources that can deliver the required momentum to the solar photosphere within a very short timescale coinciding with the start time of a hard X-ray impulse.

Recent observations, which reported helioseismic emission from the solar flares of 2003 October 28 and 29 using the helioseismic holography technique (Donea and Lindsey, 2005; DL05 thereafter), revealed another 5 sources occurred in the active region NOAA 10486 with the seismic emission at frequencies from 3 mHz to 7 mHz, 4 of them in the flare 28 October 2003 (DL05). The flare hard X-ray emission at the very start at 11:01:00 UT was observed by KORONAS \cite{kuea06} and INTEGRAL \cite{grea04,taea05} while the RHESSI started observations only at 11:06:00 UT \cite{hukr06}. However, two of the 4 sources for this flare reported by DL05 were well  aligned with the hard X-ray and $\gamma$-ray signatures observed by RHESSI after 11:06:00 UT \cite{hukr06}.  

Hence, in order to establish a connection between high energy particles and the seismic source agents, let us investigate in more detail the velocities of vertical and horizontal displacements, or the ridges, associated with these seismic waves in the flare 28 October 2003, by applying the time-distance diagram technique (TD-method, thereafter) (Kosovichev and Zharkova, 1998) to the MDI dopplergrams and by deducing the momenta required to cause the observed ridges. Then we can compare them with those delivered by high energy particles of different kinds via low temperature hydrodynamic shocks occurring in the chromosphere in response to the particle injections and travelling towards the photosphere.  

The observations used in this study are presented in \S~\ref{data} describing the active region morphology in \S~\ref{armagn}, the available high energy observations in \S~\ref{hxr}, the time-distance diagrams technique in \S~\ref{td_theory} and the observed ridges of the seismic sources detected in \S~\ref{td_obs}. A theoretical basis for the energy transport from the corona to the lower atmosphere is described in \S~\ref{theory} including the heating functions of different particles in \S~\ref{heat_f} and a hydrodynamic response to the injection of particle beams or jets in \S~\ref{hydro}. The evaluation of the wave parameters in the helioseismic sources and their comparison with the momenta delivered by hydrodynamic shocks for different kinds of particles is discussed in \S~\ref{results}. The conclusions are drawn in \S~\ref{cons}.

\section{Description of the observations} \label{data}
\subsection{Active region morphology and magnetic field}\label{armagn}
The X17.2 flare occurred on the 28th October 2003 in the very active region NOAA 10486 at the location 18E20S. It started as observed by GOES from 9:41 UT lasting until 11:24 UT with the maximum at 11:10 UT in soft X-rays and until 18:00 UT in $H_\alpha$ emission.  The active regions  NOAA 10486 had a complex delta-sunspot and produced dramatic flare activities in the descending phase of the solar cycle 23 with 3 X-class flares, i.e. an X17 flare on 2003 October 28, an X10 flare on 2003 October 29, and an X8.3 flare on 2003 November 2 and many weaker ones \cite{yaea05}. 

By tracing the changes of the sunspot group simultaneously with the TRACE white-light images, the penumbral segments are found to decay rapidly and permanently right after each of three X-class solar flares occurred in this region with the neighbouring umbral cores becoming darker \cite{liea05}. These variations are concluded to reflect the changes of the photospheric magnetic fields associated with the decaying penumbral areas  with some parts of them being converted into the umbral field. This implies the emergence of a new magnetic flux along the magnetic neutral line and a strong magnetic shear developed in this active region that plays an important role as the trigger of the X-class flare on October 28 \cite{yaea05, ishi04}.  

\begin{figure} 
\includegraphics[scale=0.9]{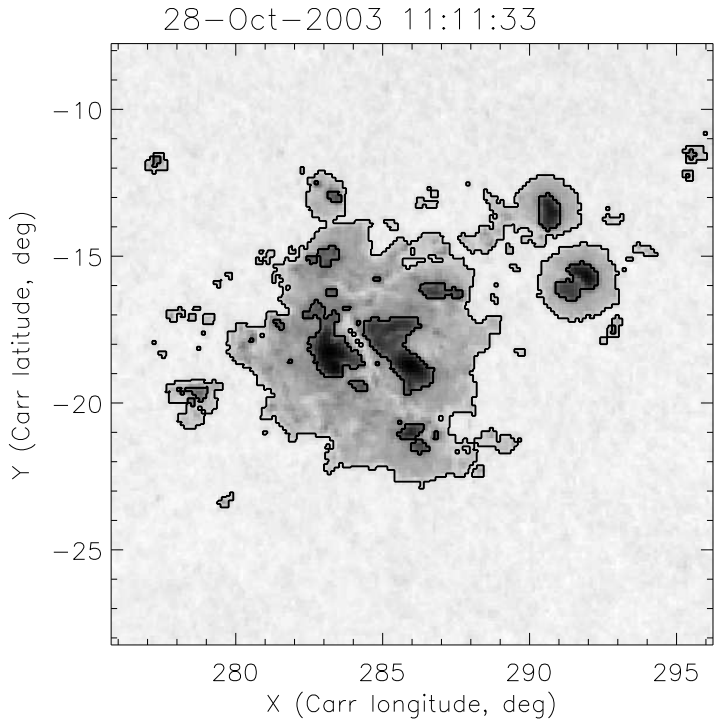}
\caption{The sunspot group located in the active region NOAA 10486 with the large trailing and 2 smaller leading sunspots obtained in the white light image at 11:05:11 UT from the Solar Feature Catalogue.}
\label{spots}
\end{figure}

\begin{figure}
\includegraphics[scale=0.80]{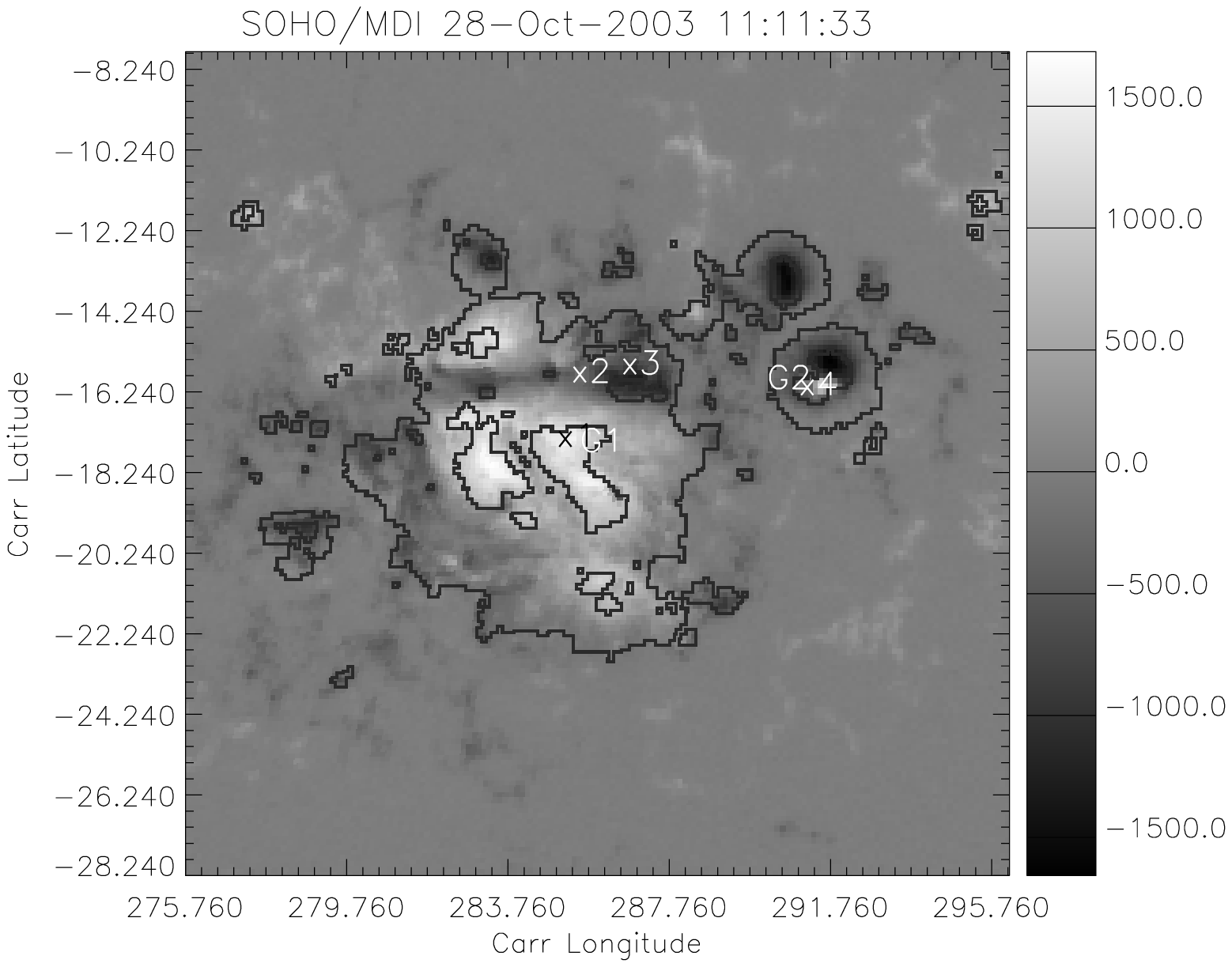}
\includegraphics[scale=0.80]{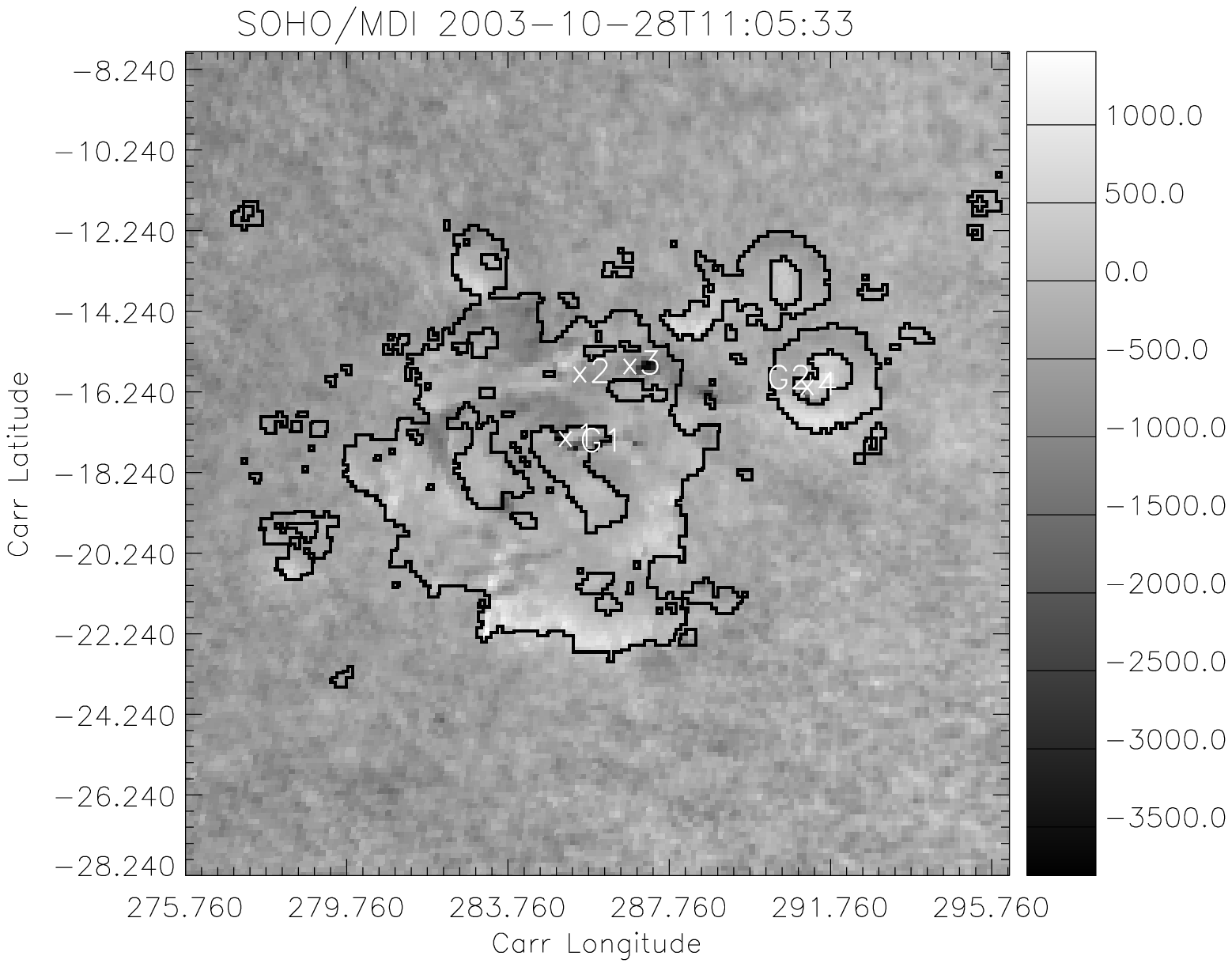} 
\caption{The same sunspot group as in Figure 1 over plotted onto the MDI magnetogram 11:11:33 UT (upper plot) and on the dopplergram  (bottom plot) with the locations of the seismic sources marked by X1-X4 (columns 4-6 in Table 1) detected by the holographic method \cite{doli05} with the gamma-ray sources observed by RHESSI marked as G1 and G2 \cite{hukr06}. }
\label{dop_mag}
\end{figure}

In the present paper we use the MDI dopplergrams obtained aboard SOHO  from 11:00 UT until 12:00 UT with 1 minute cadence supported by the MDI magnetograms and white light images taken from the Solar Feature Catalogues (SFCs) at the times closest to the flare setup time \cite{zhea05}. In Figure \ref{spots} all the sunspots detected in the MDI white light image at 11:05:33UT with the automated technique \cite{zhsea05} taken from SFCs are depicted with their umbras and pores (the upper plot) and overplotted onto the dopplergrams taken at 11:05:33 UT (from Figure \ref{dop_mag})(the lower plot). It appeared that 3 out of the 4 seismic sources X1-X4 reported by DL05 occurred either around the new umbras appearing next to the old ones (X1, X3) or in a new magnetic flux of the opposite polarity appearing in the existing umbra (X4) (see Figure \ref{dop_mag}).

\subsection{Hard X-ray, $\gamma$-ray emission and accelerated particles}\label{hxr}

The flare 28 October 2003 started in high energy emission before 11:00 UT that was only observed by KORONAS \cite{kuea06} and INTEGRAL \cite{grea04,taea05} while the RHESSI payload did not start the observing until 11:06:00 UT \cite{hukr06}. There is also a very strong CME, an interplanetary shock wave with an onset time of 11:01:39 UT and high energy particles registered at the Earth orbit\cite{kuea06, miea05}. 

\begin{figure}
\includegraphics[scale=1.1]{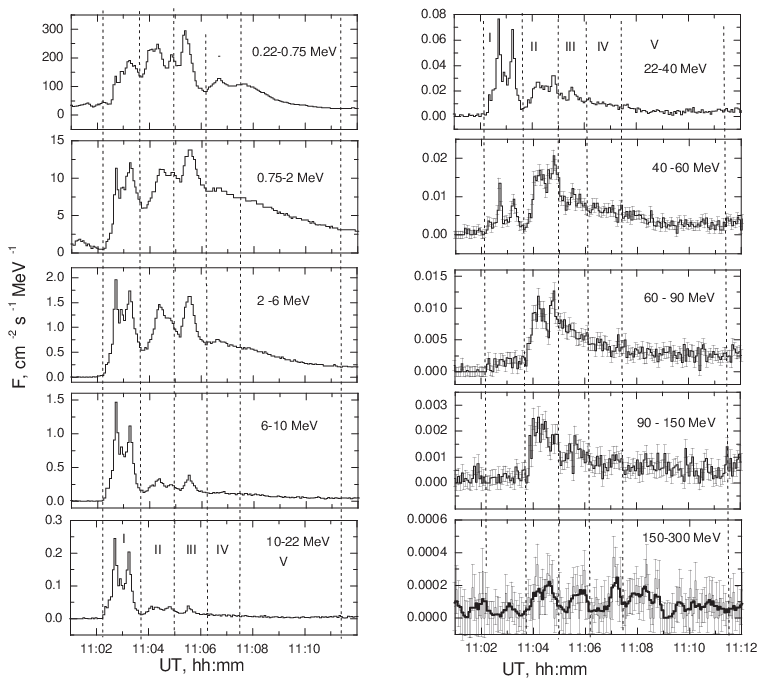} 
\includegraphics[scale=1.1]{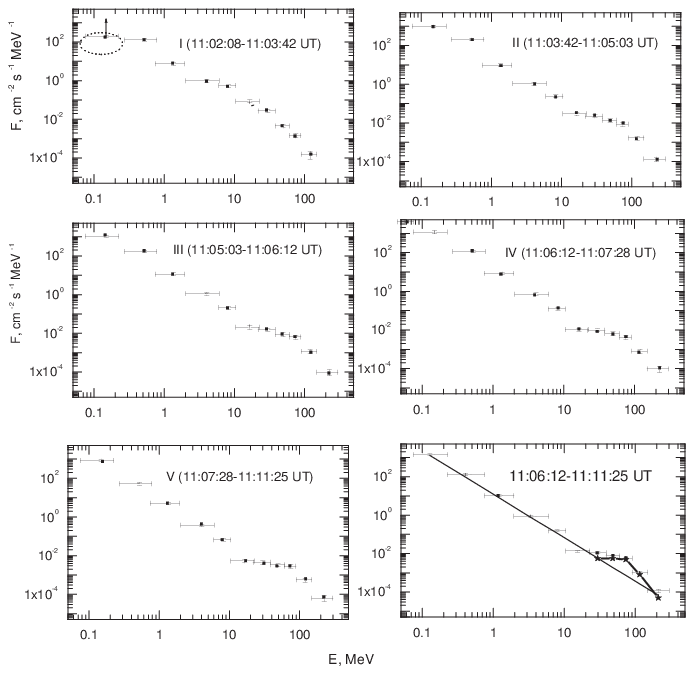} 

\caption{ The light curves (the two left columns) and the differential spectra (the righ two columns) in different energy bands obtained by the SONG (Solar Neutron and Gamma-ray) instrument aboard KORONAS during the whole flare duration (a courtesy of Dr. V.Kurt (Moscow University) and the KORONAS team \cite{kuea06}). } 
\label{coronas}
\end{figure}

The flare light curves in $\gamma$-rays measured by the SONG {\bf SO}lar {\bf N}eutrons and {\bf G}amma rays) instrument aboard KORONAS are plotted in Figure \ref{coronas}a with the particle energy spectra in Figure \ref{coronas}b (a courtesy of Dr. V.Kurt and the KORONAS team \cite{kuea06}). Similar light curves were observed in the $\gamma$-continuum (2.8-3.7 MeV and 7.6-10 MeV) and $\gamma$-lines (2.22 MeV, 4.44 MeV and 6.13 MeV) by the instruments aboard INTEGRAL satellite (Figure 5 in \cite{grea04}). 

These light curves revealed the three distinct phases in the flare evolution: a short impulsive phase A (under one minute) with a sharp increase of the continuum emission in both channels with the photons detected up to 15 MeV, the SPI energy limit, and the two longer phases B and C, contained a sharp increase of the 2.8-3.7 MeV  continuum emission (phase B) and much smoother increase of the line emission in all 3 lines (phase C).

The start time of the flare impulsive phase was about $11:01:00\div 11:02:00$ UT (see Figure \ref{coronas}) based on the measurements by INTEGRAL \cite{grea04} and KORONAS \cite{kuea06}. From 11:01:39 UT until 11:05:40 UT the brightest outburst in energy was in the range of $0.5\div 40 MeV$. While the protons $>100\, MeV$ were only detected at about 11:06:00 UT when also RHESSI started to observe. The KORONAS light curves show the 2 rather distinguished peaks in $\gamma$-ray emission at 11:02:00 UT and 11:03:00 UT in the ranges from 0.5 MeV to 41 MeV (corresponding to the phase A and the start of B in the INTEGRAL) and another 2 peaks appearing later at about 11:06:00 UT in the lower energy ranges of 0.15 - 4 MeV and 26 MeV - 100 MeV corresponding to the phase B from INTEGRAL. 

In the first phase the 2 $\gamma$-continuum peaks are concluded to be produced mainly by a bremsstrahlung spectrum generated by electrons with energies up to 150 MeV with a small $\gamma$-ray  increase in the range of 1.5-7 MeV \cite{grea04, kuea06}. In this phase the $\gamma$ line emission was also observed by the KORONAS that indicates a presence of protons with the energies $>30\, MeV$ but not higher than 200 MeV because of the absence of photons from the $\pi$-decay process. The photon energy spectra obtained during the first phase (see Figure \ref{coronas}b) are single power laws with the spectral indices about 2 at the lower energy part ($<70$ KeV) and 3.5-4 at higher energies \cite{kuea06,kurt06}. 

The second phase, a delayed one,  where the other 2 peaks are observed in $\gamma$-rays by KORONAS, INTEGRAL and RHESSI, has very noticeable plateaus in the energy spectra in the range of 25-100 MeV (see the middle and bottom rows in Figure \ref{coronas}b) first appearing at about 11:06:10 UT. Also there are higher energy protons $>200 MeV$ appearing after 11:06:00 UT indicated by a presence of $\pi$-decay photons \cite{kuea06,shea04}.

The images of the sources of hard X-ray (200-300keV) and $\gamma$-ray emission (2.2MeV) obtained by RHESSI \cite{hukr06} are marked as G1 and G2 on the active region image in Figure \ref{dop_mag}. The images reveal that at least after 11:06:00 UT when RHESSI started to observe there were 2 footpoints with the hard X-ray and $\gamma$-ray sources, which have slightly different spatial locations with the seismic sources located between them. 

The spectral indices of the proton energy spectra deduced from the ratios of $^{12}C$ and $^{16}O$ lines observed in the phase B, or after 11:06:00 UT, from INTEGRAL vary from 3 to 3.8 \cite{taea05}. This is close to those of $2.8\mp 0.4$ reported also after 11:06:00 UT by the RHESSI measurements from the de-excitation line 2.22 MeV and positron annihilation line 511 keV with a total number of protons of about $10^{33}$ \cite{shea04}.

Hence, in this flare there are ample indications from the observations of high energy emission about a few events with particles arriving at various times, starting 11:02:00 UT until 11:07:00 UT at the footpoints. 

\subsection{The time-distance diagrams technique } \label{td_theory}

Now let us investigate from the MDI dopplergrams this flare evolution at the photospheric level and below. We use the one minute cadence dopplergrams for the hour starting from 11:00:00 UT. Since the typical oscillation frequencies associated with quakes \cite{kozh98, doli05} are higher than the background oscillations (3mHz), we also apply frequency filtering centred at 5-6 mHz in order to increase the signal-to-noise ratio.

The obtained velocity distributions in the area with radius of 120 Mm are fit by a circular wave using the angular Fourier transform for the angle $\theta$ that can be described by the angular Fourier transform \cite{chri03} for the angle $\theta$ as follows:
\begin{equation}
v(r,\theta,t)= \Sigma _{m=0,2} v^0_{m} (r,t) e^{im\theta},
\end{equation} 

where $m=0$ denotes a circular wave, $m=1$ - a dipolar wave and $m=2$ a a quadruple wave. Practically, as the previous observation shown \cite{kozh98}, it is not expected to observe the waves higher than these three types because of the observational noise. In the present study we extract only a circular wave, while the dipolar wave was also registered for this flare by Kosovichev (2006).

The MDI dopplergrams are re-mapped into polar coordinates with the centres around the location of the holographic seismic sources X1-X4 \cite{doli05}. The new velocity distributions $v(r,\theta,t)$ are obtained as the function of time $t$ and their distance $r$ from the centre (defined with the accuracy of a single pixel).

 Then we extract the velocities $v^0_0(r,t)$ for every $r$ averaged over various angles $\theta$ for $m=1$ in the datacube of 120 Mm and measure for the different times the horizontal displacements of the propagating seismic waves \cite{chri03}, i.e. $v^1_0(r_1,t_1)$, $v^1_0(r_2,t_2)$ that can be plotted for different times (axis Y) and distances (axis X) as a time-distance diagram \cite{kozh98}.

\subsection{The observed seismic sources} \label{td_obs}

In order to precisely detect the seismic wave centres, we selected the areas of 20x20 pixels around the locations provided by DL05 (Table 1, the last three columns). Then for their 4 sources (X1-X4) from Table 1 we obtained 400 TD-diagrams, which were visually investigated for the pixel locations with detectable ridges denoting the seismic wave propagation. Then the pixels with the detectable ridges were used to define the total areas for each quake, and their centres of gravity were used as the centres of the seismic waves.

\begin{figure} 
\centerline{\includegraphics[scale=0.8]{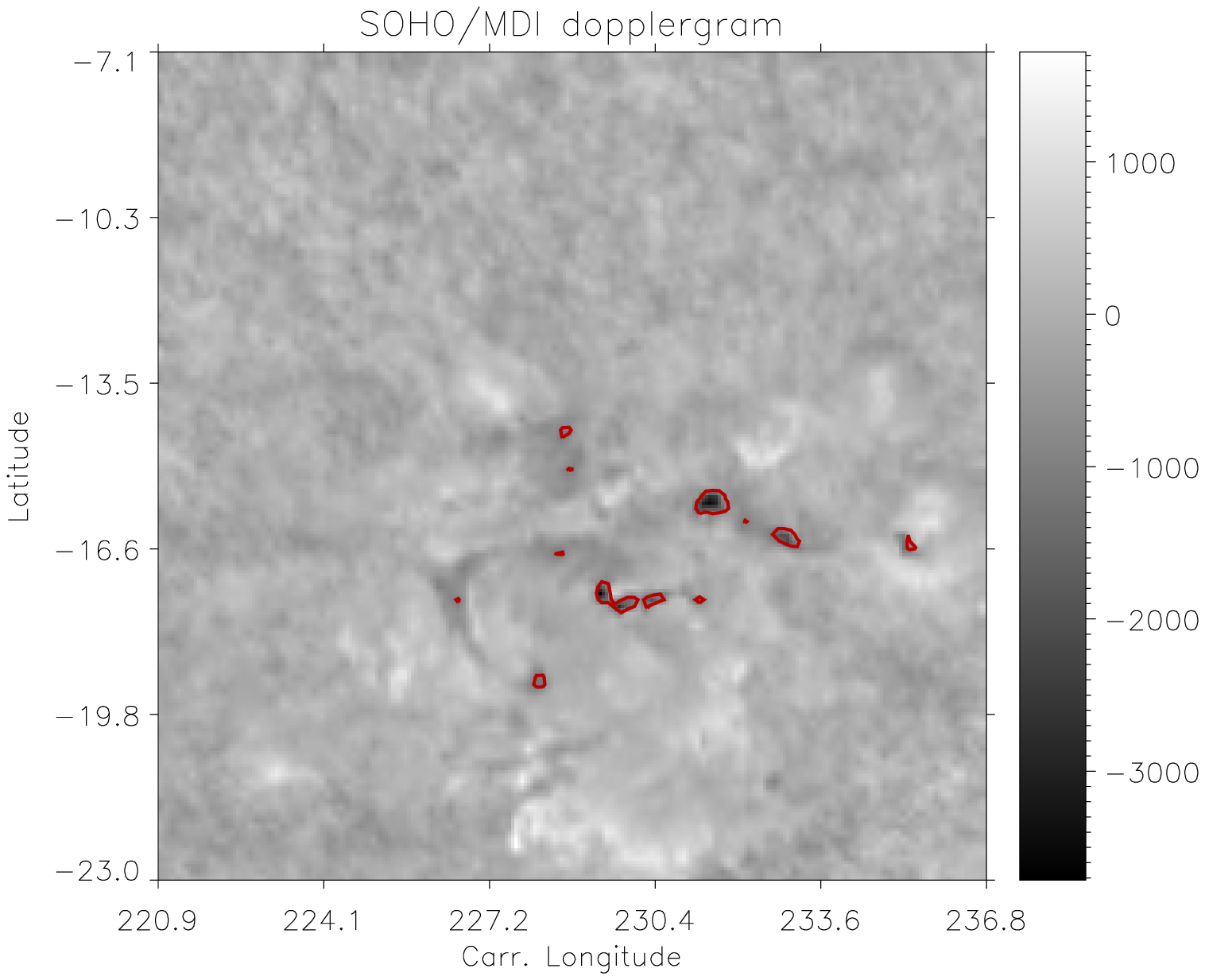}} 
\includegraphics[scale=0.8]{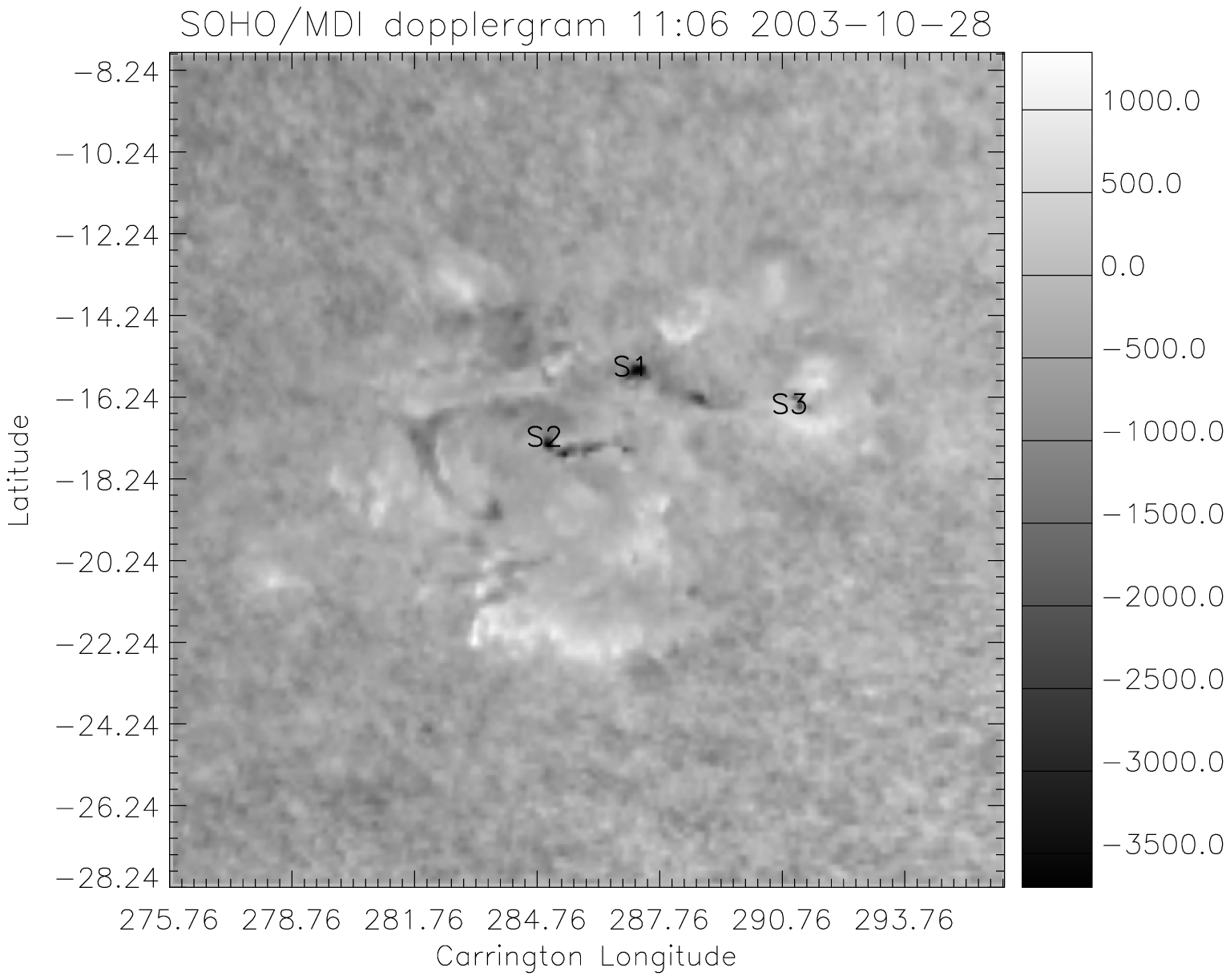} 
\caption{ The locations of 11 Doppler sources (red contours) with the downward motion higher than 1 km/s (top image) but only 3 of them with the detectable seismic responses in the locations specified in columns 1-3 of Table 1 (bottom image).} 
\label{dop11}
\end{figure}

For the detection of seismic sources we use the one minute cadence dopplergrams starting from 11:00:00 UT until 13:00:00 UT. In general, in all the MDI dopplergrams we have detected 11 locations with downward motion larger than 1 km/s (Figure \ref{dop11}) using the automated technique (Zharkov et al., 2005), while only three of them, S1, S2 and S3, have revealed detectable ridges, or quakes (Figure \ref{sources}). For each seismic source we extracted the datacube of 120x120 Mm centred in the centre of gravity of the downward Doppler motions in this source, re-mapped them into the polar coordinates and applied to the Fourier technique (formula 1) for a distance from the centre where the initial impulse is applied (the center of gravity) up to 120 Mm. The deduced locations for the seismic sources S1, S2 and S3 with the detectable ridges are summarised in the Table 1 (columns 1-3) compared with the sources X1-X4 detected by DL05 (columns 4-6). 

The  downward velocities $V_{vert}$ (Figure \ref{vel_vert}) directly measured from the dopplergrams in these 3 sources were 2.15 km/s for S1, 2.0 km/s for S2 and 1.75 km/s for S3, while for another 8 of 11 sources the downward velocities were lower than 1.6 km/s . The durations of the downward motions do not exceed 1.5 -2.0 minutes, starting at or after 11:05:00 UT with the maximum a minute later and then decreasing for another 30-60 seconds back to the pre-flare magnitude. 

The time-distance diagrams for our sources S1-S3 are presented in Figure \ref{sources} without and with the theoretical ray paths (white solid lines) for the source S1 (upper plots), S2 (middle plots) and S3 (bottom plots). The start times of the seismic waves are slightly different in each source varying from about 11:05:00-11:06:00 UT for the source S1 to 11:06:00 UT for the sources S2 and S3. These times are  more than 3 minutes later than the first maximum of hard X-rays for the source S1 but close to the second maximum in both HXR and $\gamma$-rays pointing  to the presence of high energy protons ($>200$ MeV) for the sources S2, S3 \cite{kuea06,shea04}. 

The source S1 (Figure \ref{sources}, upper plots), corresponding to the source X3 (DL05), at the very start of the seismic response, or ridge, reveals the initial horizontal velocity of about 42 km/s that ncreased in 50 minutes up to 140 km/s at the distances of 120 Mm.  The ridge in the source S2 (middle plots), or the source X1 in DL05, was slightly steeper compared to the source S1, i.e. the seismic wave was slower, its velocity varies from 38 to 128 km/s in 53 minutes. The ridge in the source S3 (Figure \ref{sources}, the bottom plots), the source X4 in DL05, shows the lowest velocity variations from 34 to 114 km/s in 60 minutes. Hence, the initial velocity in the source S1 is higher and the seismic wave propagates towards the 120 Mm edge of the datacube slightly faster than in other two sources S2 and S3.

The areas of the seismic sources, defined by a presence of the downward motions and by detectable ridges, were about $5.05\times 10^{17}$ $cm^2$ for the source S1, $3.34\times 10^{17}$ $cm^2$ for the source S2 and $3.22\times 10^{17}$ $cm^2$ for the source S3. Then by using the downward velocities above (see Table 2) and comparing with the theoretical seismic ridges (Kosovichev and Zharkova, 1995, 1998) one can deduce the momenta required to cause the observed ridges: $4.0\times 10^{22}$ $g\cdot cm\cdot s^{-1}$ (S1), $3.7\times 10^{22}$ $g\cdot cm\cdot s^{-1}$ (S2) and $3.1\times 10^{22}$ $g\cdot cm\cdot s^{-1}$ (S3). 

The locations of the seismic sources S1, S2 and S3 found from the TD diagrams are slightly different than those reported from the holographic method DL05 that can be a result of different sensitivity of the techniques applied by us and DL05. The TD technique does not produce distinguished ridges for the source X2, which was seen after 11:07:00 UT (DL05). There was a weak downward source in the location of the source X2 observed by MDI between the times 11:02:00 and 11:03:00 UT (the first maximum in hard X-rays) as was spotted in Figure 5 by DL05. While the locations of the sources S1-S3 coincide with the dark spots inside the sunspot umbras, there is no indication of any new umbras in the location of the source X2. 

The absence of the 4-th source in the TD diagrams can have a number of explanations. The source X2 can be rather weak and, thus, cannot be detected by the TD technique while seen by the holographic method. Other options are either that this source is an interference of the seismic waves produced by the two sources (possibly, S1 and S2) or it is located so closely to the source S1 that we merged them into the extended S1 source. However, the problem with the latter is that the TD diagram for S1 shows its start time at 11:05:00 UT. These are very puzzling questions that we with the authors of the holographic method (DL05) are planning to investigate in the future for this and other flares. 

\section{The particle energy transport} \label{theory}

In order to evaluate the transfer into the ambient plasma of the particle momenta and energy after their injection from the top into a flaring atmosphere, let us investigate their heating functions and resulting hydrodynamic responses. 

Let us assume that the protons or electrons accelerated in a reconnecting current sheet (RCS) with a strong longitudinal magnetic field occurred during this flare then they can be ejected as power law beams and completely or partially separated into the opposite footpoints of the same loop \cite{zhgo04,zhgo05a}. In addition to the electron or proton beams, let as consider the particles with Maxwellian (thermal) energy distributions shifted to higher energies accelerated along the separatrices of the RCS (Gordovskyy et al., 2005). 

Therefore, we consider the 4 kinds of high-energy particles whose energy losses are converted into the ambient plasma heating: fast electron beams and fast proton beams with power law energy distributions and slow electrons and protons of the separatrix jets with thermal energy distributions.

\subsection{The heating functions by high energy particles} \label{heat_f} 
 
The energy deposition functions, or heating rates, for these particles are calculated using distribution functions found from the full kinetic approach by solving the Fokker-Plank equation for electrons losing their energy in collisions and Ohmic heating (Zharkova and Gordovskyy, 2005b) and protons in the generation of kinetic Alfven waves (KAWs) and their dissipation via Cherenkov's resonance (Gordovskyy, 2005; Gordovskyy et al., 2005).

The volume heating rates by all kinds of high energy particles simultaneously present in the flaring atmosphere are calculated from the particle distribution functions as a vertical gradient of the beam energy flux:

\begin{equation}
S(\xi) = - n(\xi) \frac {d(F_e(\xi)+F_p(\xi)+F_j(\xi))}{d\xi},
\end{equation}

where $\xi=\int_{z_min}^{z_max} n(z) dz$ is a column density, i.e. a number of the ambient particles in the area of $1 cm^2$ on a line of sight from the height $z_min$ to $z_max$, $n(z)=n(\xi)$ is a total density of the ambient plasma at a given height, $F_e(\xi)$, $F_p(\xi)$ and $F_j(\xi)$ are the energy fluxes carried by fast electrons, "fast" protons and "slow" protons (of separatrix jets), respectively. We do not include the energy losses by slow electrons since these are negligable compared to the other particles (Gordovskyy et al., 2005). The heating rate per particle of the ambient plasma $P(\xi)$ is related to the volume heating rate $S(\xi)$ as $P(\xi) = \frac {S(\xi)}{n(\xi)}$. The variations of the density are considered from a hydrodynamic response below.

 \subsection{Hydrodynamic response to the particle injection} \label{hydro}

Let us now consider a hydrodynamic response of the 1D solar atmosphere to the injection of electrons and/or protons by taking into account the continuity, momentum and energy equations for the ambient electrons and protons/ions. 

The physical conditions in a flaring atmosphere can be described by a plasma density $n$; electron $T_e$ and ion $T_i$ temperatures and a vertical velocity $v$. All these parameters vary with a vertical coordinate $z$, or a column density $\xi$, and time $t$. The ambient plasma response to the injection of high energy particles is described by the hydrodynamic equations (see e.g. Somov et al., 1981; Fisher et al., 1985a-c):\\

a) Continuity equation:

\begin{equation}
\frac{\partial n}{\partial t}\,+\,n^2\frac{\partial v}{\partial \xi}\,=\,0,
\end{equation}

b) Momentum equation:

\begin{equation}
\frac{\partial v}{\partial t}\,+\,\frac{1}{\mu}\frac{\partial }{\partial \xi}[nk_B(T_i+xT_e)]=4/3\frac{1}{\mu} \frac{\partial }{\partial \xi}[\eta _i n\frac{\partial
v}{\partial \xi}]\,+\,g_{\odot},
\end{equation}

c) Energy equation for ions:

\begin{equation}
\frac{nk_B}{\gamma-1}\frac{\partial T_i}{\partial t}\,-\,k_BT_i\frac{\partial n}{\partial t}=
\frac 4 3 \eta _i n^2 (\frac{\partial v}{\partial \xi})^2\,+\,Q(n, T_e,T_i),
\end{equation}

d) Energy equation for electrons:

\begin{equation}
\frac{nk_B}{\gamma-1}\frac{\partial (xT_e)}{\partial t}\,-\,xk_BT_e\frac{\partial n}{\partial t}\,+\,n\chi \frac{\partial x}{\partial t}\,=
n \frac{\partial }{\partial \xi}(\kappa n\frac{\partial T_e}{\partial \xi})\,+\,P(n,\xi)\,-\,L(n,T_e) ,-\,Q(n,T_e,T_i).
\end{equation}

Here $T_i$ and $T_e$ are the ion and electron temperatures, respectively, $n$ is the ambient plasma density, $x$ is the ionization degree of the ambient plasma, $\gamma$ is the adiabatic constant, $k_B$ is the Boltzmann constant, $\kappa $ is the thermal conductivity, $\eta _i$ is the ion viscosity, $\chi$ is the full ionization energy of a hydrogen atom, $\mu=1.44m_{H}$, $g_{\odot}$ is the acceleration due to gravity of the Sun. $P(n,\xi)$ indicate the volume heating rates provided by electrons owing to collisions, by electrons owing to Ohmic losses, by protons owing to collisions and by KAWs, $L_{rad}$ is the volume radiative energy loss rate and, finally, $Q(n, T_e,T_i)$ is the rate of energy exchange between ions and electrons.

The solution is sought in a limited region of the solar atmosphere $\xi_{min}\leq \xi \leq \xi_{max}$ with the minimum boundary located at $\xi_{min}=2\times10^{17}$ $cm^{-2}$ and the maximum boundary is located deep in the photosphere at $\xi_{max}=2\times10^{24}$ $cm^{-2}$. The initial atmosphere is assumed to be in hydrostatic equilibrium, i.e. $v(0,\xi)=0$ and isothermal, i.e. $T_e(0,\xi)=T_i(0,\xi)=T_0$ where $T_0=6700^\circ $ K. The ionization degree $x$ is defined by a modified Saha formula (Somov et al., 1981). We also take into account the initial momenta delivered by the particles at injection \cite{brcr84} by assuming $v(0,\xi_{min})= \frac{P_B} m$ where $P_B$ is the sum of the momenta of all injected particles.

The initial distribution of a plasma density is defined as follows:

\begin{equation}
n(0,\xi)=n_{min}+h_0^{-1}(\xi-\xi_{min}),
\end{equation}

where $n_{min}=10^{10}$ $cm^3$ and $h_0$ is the height scale:

\begin{equation}
h_0\,=\,\frac{k_B[1+x(T_0)]T_0}{\mu g_{\odot}}.
\end{equation}
 
The radiative losses rate $L(n,T_e)$ is described by the analytical expression:

\begin{equation}
L_{n, T_e}=n^2 x L(T_e) + nL_H(n,T), {\rm (erg/cm^3/s)} 
\end{equation}

where the radiative loss function $L(T_e)$ is taken for the coronal abundances of elements in optically thin plasma \cite{cotu71} and $L_H(n,T_e)$ are the radiative losses in all hydrogen lines calculated for the optically thick atmosphere \cite{zhko93}. 

The boundary conditions are defined as follows. 1). We assumed that there is no an initial heat flux on the top boundary, i.e. $\frac{\partial T_e(0,\xi_{min})}{\partial \xi}=\frac{\partial T_i(0, \xi_{min})}{\partial \xi}=0$; 2) the upper boundary is a free surface in the presence of the coronal pressure, i.e.
\begin{equation}
\frac{\partial v(t, \xi_{min})}{\partial \xi}\,=\,\frac{4}{3}\frac{1}{n\eta _i}[p(t,\xi_{min})-p_{cor}(z(t,\xi_{min}))],
\end{equation}
where $p(t, \xi_{min})=nk_B(T_i+xT_e$) and $p_{cor}(z(0,\xi_{min}))=n_{min}k_B[1+x(T_0)]T_0$ where the ionization degree $x$ is defined by a modified Saha formula (Somov et al., 1981).
\subsection{The momenta delivered by beams and hydrodynamic shocks} \label{momenta}
\subsubsection{The momentum delivered by a proton and/or electron beam}\label{pmom}

The momentum $P_{e,p}$ delivered in pure collisions by an electron or proton beam with a spectral index $\gamma_{e,p}$ and a lower energy cutoff $E_{low_{e,p}}$ can be evaluated as \cite{zhgo05b}:

\begin{equation} 
P_{e,p}\,=\,\sqrt{2 m_{e,p}} K \,\int_{E_{low_{e,p}}}^{\infty} E^{-\gamma_{e,p}+0.5} dE, 
\end{equation} 

where $m_{e,p}$ is the electron or proton mass, $E_1$ is the lowest energy in the relevant particle spectrum (theoretical lower energy cutoff) and K is the normalisation  constant that can be found from the total number of measured particles $N_{e,p}$: 
 
\begin{equation} 
N_{e,p}\,=\, K\,\int_{E_{low_{e,p}}}^{\infty}\, E^{-\gamma_{e,p}}\,dE\,= \frac K \gamma E_{low_{e,p}}^{-\gamma\,+\,1}, 
\end{equation} 

where $E_{low_{e,p}}$ is the measured lowest energy of electrons or protons. 
Hence, without taking into account pitch angle scattering, the momentum delivered by a proton beam can be evaluated by substituting the constant $K$ found from $N_p$ into the equation for $P_{e,p}$ and performing the integration. This will result in the following: 
\begin{equation} 
P_{e,p}\,\simeq \,\sqrt{2m_p}\,N_{e,p} \,\frac{\gamma_{e,p}}{(\gamma_{e,p}+0.5)}\,\frac{E_1^{-(\gamma_{e,p}-0.5)}}{E_{low_{e,p}}^{-(\gamma_{e,p}-1)}}. 
\end{equation} 
If we assume that $E_1=E_{low_{e,p}}$, then the momentum delivered by electrons/protons without pitch angle scattering can be evaluated as: 

\begin{equation} 
P_{e,p}\,\simeq \,\sqrt{2m_{e,p}}\,N_{e,p} \,\frac{\gamma_{e,p}}{\gamma_{e,p}+0.5}\cdot E_{low_{e,p}}^{-0.5}. 
\end{equation}

It should be emphasized that this is the upper limit of the momentum carried downwards to the photosphere by electrons or protons since it is calculated without taking into account pitch angle scattering and wave dissipation for protons or Ohmic dissipation for electrons that can reduce its magnitude by a few factors \cite{goea05}. However, it allows the comparison of the momenta delivered by high energy particles with those measured from the downward motions in dopplergrams and from the TD diagrams.

\subsubsection{The momentum delivered by a hydrodynamic shock} \label{hd_mom} 

Let us also evaluate the momentum delivered by a hydrodynamic shock using the simple formula: 

\begin{equation}    
P_{hd}\, =\, \Sigma _t mv(t)    
\end{equation}    

where the summation is done over the time from 0 to $\tau$ where $\tau $ is a duration of the impact causing the seismic waves, $m $ is the mass  of the plasma delivering the momentum related to the flaring area $A$ where the momentum is deposited, $V$ is a starting velocity  of the plasma at the moment of impact and $t$ is a duration of the impact. 

For the known plasma mass density $\rho= m_H\cdot n$ where $n$ is the particle density per volume defined from hydrodynamic solutions, this  equation can be re-written as  follows: 

\begin{equation}    
P_{hd}\, =\,\Sigma _t mv(t) \approx \rho \cdot A\cdot v^2\cdot\tau,     
\end{equation}    
 
where $\rho $ is an average density of the plasma delivering the momentum, $A$ is the flaring area where the momentum is deposited, $v$ is an averaged velocity of the plasma propagation at the moment of impact and $\tau $ is the duration time of the impact. 

\section{Results and discussion } \label{results}

\subsection{The momenta delivered by beams} \label{est_mom}

Let us try to establish the agents delivering the momenta reported in Table 2 (\S~\ref{td_obs}) by investigating the parameters of high energy particles associated with each source from the hard X-ray  and $\gamma$-ray observations by CORONAS, INTEGRAL and RHESSI in \S~\ref{hxr}. 

{\it The source S1}

For the source S1 one has no $\gamma$-ray emission but only hard X-ray photon differential spectra (or mean flux) as presented in Figure \ref{coronas}b with the upper energy of 150 MeV and the lower energy cutoff about 18 keV (the left upper plot for $11:02:00\div 11:03:00$ UT, Kurt, 2006). The photon spectral index in this plot was about $\delta_{high} \,=\,3.5-4.0$ at the energy range from 70 keV to 60 MeV and $\delta_{low}\,=\,1.5$ for the energy range of $10\div 70$ keV (Figure \ref{coronas}). Since this was a very strong flare, we assume that the precipitating beam has induced a very strong electric field and its hard X-ray emission was dominated by the Ohmic energy losses \cite{zhgo05b,zhgo06}.  This assumes the spectral index $\gamma$ of a precipitating electron beam has to be nearly the same as the index $\delta_{high}$ of the photon spectrum at higher energy. 

The difference between the spectral indices $\delta_{high}$ and $\delta_{low}$ can provide us with the beam initial energy flux $F_0$ for the selected spectral index of an electron beam, i.e. for the beams with $\gamma$ varying from 3.5 to 5, the difference in the photon indecies varies from 2.0 to 2.5. Hence, from Figure 12 \cite{zhgo05b} we can deduce the initial energy flux of beam electrons that can vary from $1\cdot 10^{12}$ $erg/cm^2/s$ for $\gamma=3.5$ to $4\cdot 10^{11}$ $erg/cm^2/s$ for $\gamma=5$ \cite{zhgo05b}. These fluxes can be carried out by electron beams with the initial densities of $1\cdot 10^{9}$ $cm^{-3}$ and $6\cdot 10^{8}$ $cm^{-3}$, respectively. 

Let us calculate the momenta delivered  to the photosphere by electron beams with such parameters using the technique described in \S~\ref{pmom}. For the flare area of about $5.05\times 10^{17}$ $cm^2$ defined by the area of a downward Doppler motion Figure \ref{dop11} in the source S1, the momentum delivered to this area by the electron beam was about $2 \cdot 10^{20}$ $g \cdot cm /s$. Obviously, this is not sufficient to deliver the required momentum of $(3-4)\times10^{22}$ as reported in Table 2 and to directly cause the ridge observed in the TD diagram for S1 (see \S~\ref{td_obs}).

{\it The sources S2 and S3}.

The sources S2 and S3 appear close to the locations of hard X-ray emission and within the circle denoting $\gamma$-ray emission observed by RHESSI \cite{hukr06}. As it was noted in \S~\ref{hxr} the spectral indices of the proton energy spectra observed by INTEGRAL in the phase B after 11:06:00 UT (from the ratios of $^{12}C$ and $^{16}O$ lines) vary from 3 to 3.8 \cite{taea05} or by the RHESSI measurements (from the de-excitation line 2.22 MeV and positron annihilation line 511 keV) vary $2.8\mp 0.4$ with a total number of protons observed estimated at about $10^{33}$ \cite{shea04} and the lowest energy about $30 MeV$, or $\approx 4.8\cdot 10^{-5}$ erg. However, this energy can be decreased to 2 MeV without affecting the 2.22 MeV and annihilation line emission \cite{shea04}.

Then the momentum delivered by such a proton beam to the chromosphere where the $\gamma$-emission is measured, can be about $\sim 2.2\cdot 10^{22}$ $g \cdot cm/s$ for the lower energy of 30 MeV and according to equations (12) and (14), increases as $(2 MeV/30 MeV)^{-\gamma +0.5}\times\frac{\gamma}{\gamma+0.5}$ to $\sim 5.2\cdot 10^{24}$ $g \cdot cm/s$ for the lower energy of 2.0 MeV and $\gamma= 3.0$. This range superbly covers those momenta derived in the sources S2 and S3 (Table 2). Hence, protons can be the agents in these two sources delivering the sufficient momenta to the region where MDI measures the Doppler velocities. Of cource, as pointed in \S~\ref{pmom}, the momentum evaluated from formulae (12-14) provide the upper limit since other proton or electrons scattering mechanisms can slightly reduce a number of particles reaching a given depth from the top depending on beam parameters. However, for moderate hard beams, as reported for this flare (spectral index of 3.3), this difference is not very noticeable (Zharkova and Gordovskyy, 2005b).
\subsection{Simulated heating functions }\label{heat_dis}
Now let us compare the heating rates of the three kinds of particles considered in \S~\ref{theory} and \S~\ref{heat_f}. The timescale within which particle of each kind can reach the photosphere is about 1 s for electron beams, 2-5 s for proton beams and 10-20 s for protons of separatrix jets for the standard loop length of $10^9$ cm (Zharkova and Gordovskyy, 2004; Gordovskyy et al. 2005). Therefore, the propagation time for each kind of particles is short enough to contribute to the ambient plasma heating and to form a shock, or a lower-temperature condensation, appearing as a result of the hydrodynamic response to these particle injection.

The heating rates simulated from the full kinetic approach are presented in Figure \ref{heat_g}: curve A - for an electron beam with the initial energy flux $F_0 = 1.4\cdot 10^8 erg/cm^2/s$, spectral index $\gamma=2$ and the lower energy cut-off $16 keV$, curve B - for the separatrix jet protons with the  initial energies $E_0=1 MeV$ and initial energy flux of $4\cdot 10^{11}erg/cm^2/s$ , curve C - for an electron beam with the initial energy flux $F_0 = 10^{10} erg/cm^2/s$, spectral index $\gamma=5$ and the lower energy cut-off $16\,keV$, curve D - for a proton beam with the  initial energy flux $F_0 = 4\cdot 10^{10} erg/cm^2/s$, spectral index $\gamma=1.5$ and the lower energy cut-off $40\,keV$ and curve B for the separatrix jet protons with the initial energies $E_0=1 MeV$ and initial energy flux of $4\cdot 10^{11}erg/cm^2/s$.  

It can be seen that the heating by electron or proton beams with power law energy distributions is strongly dependent on the initial beam parameters: softer and weaker beams deposit their energy mainly in the corona and upper chromosphere while harder and more powerful beams deposit more energy deeper in the lower chromosphere (compare the curves A and C for hard and soft electron beams and D for a soft proton beam) (Zharkova an Gordovskyy, 2005b, Gordovskyy et al., 2005).  

Electron beams are considered to deposit their energy more evenly in depth compared to the proton ones (compare the curves A,C with D). Proton  beams deposit the bulk of their energy via generation of KAWs with their following dissipation in Cherenkov's resonance at the flaring atmosphere depths where their velocities are higher than the local Alfven ones (Gordovskyy et al., 2005). It can be noted that the heating by KAWs induced by protons has the two regions where this condition stands: in the upper corona because of their initial exponential distributions (the first curve B) and at the lower chromosphere because of reduction of the local Alfven speed and of the proton exponential distributions (the second, spike-like curve B).

The heating of the upper atmosphere by proton beams can be even more noticeable after Coulomb collisional losses are taken into account which will make the proton distributions in the chromosphere even more exponential. While strongly affecting the heating by proton or electron beams of the corona and upper chromosphere (before the column depths of $10^{20}$ $cm^{-2}$), the collisions do not change significantly the heating of the lower chromosphere  where the Cherenkov resonance is dominant (Gordovskyy et al., 2005). Another heating mechanism considered for beam electrons is Ohmic dissipation in a self-induced electric field that contributes significantly to the heating of the coronal levels but again has not affected the lower atmosphere heating (Zharkova and Gordovskyy, 2005b).

The effect of these heating functions on the hydrodynamic solutions is discussed below. We include into the heating functions: collisional losses by both electron and proton beams, Ohmic losses only for an electron beam and Cherenkov resonance for thermal-like protons. Two heating functions are considered: by a pure electron beam and by a proton beam combined with quasi-thermal jet protons.

\subsection{Simulated hydrodynamic responses} \label{hd_mod}
In order to maximise a deposition at lower atmospheric levels let us compare the hydrodynamic responses caused by a hard electron beam with $\gamma=3.5$  and the intial energy flux of $~10^{11}$ $erg/cm^2/s$ (that is higher and harder than measured) with those by proton beams/jets with the spectral indices of about 3 and the initial energy fluxes of $10^{12}$ $erg/cm^2/s$ as deduced from the hard X-rays and $\gamma$-rays in \S~\ref{hxr}. 

We simulate the hydrodynamics of a two-temperature plasma heated by either electrons or protons (power laws plus those with Maxwellian energy distributions) by solving the two energy equations (for electrons and protons), continuity and momentum conservation equations and include the radiative cooling as described in \S~\ref{hydro}. 

The variations of temperature, density and macro-velocity simulated for the hydrodynamic responses are plotted for the electron beam with the parameters derived from hard X-rays (Figure \ref{hydr}, the left plots) or for the mixed proton/jet beam with the parameters derived from the $\gamma$-ray emission (Figure \ref{hydr}, the right plots). 

The presented hydrodynamic results caused by beam electrons agree with those by Somov et al. (1981), Nagai and Emslie (1984) and Fisher et al. (1985a-c). For the intense hard beam as deduced from the hard X-ray data from KORONAS (Kuznetsov et al., 2006) there is a noticeable decrease (rarification) of the total plasma density and a strong temperature increase in the corona accompanied by explosive evaporation of the chromospheric plasma into the corona (macromotion upwards) occurring in response to the beam injection. 

Starting from the column density $2\times10^{19}$ $cm^{_2}$, a collisional stopping depth for lower energy electrons of 12 keV accepted in our simulations, a low temperature condensation is formed moving as a shock with velocities of about 100-200 km/s, which are higher than the local sound velocity (Somov et al., 1981, Fisher et al., 1985). However, such a shock produced even by a powerful electron beam with the spectral index of about 3 (the lower plots) appears rather high in the upper chromosphere between the column depths of $2\times10^{19}\div 2\times 10^{20}$ $cm^{-2}$ as can be seen in Figure \ref{hydr} (the left plots). 

These two motions of the ambient plasma (upwards and downwards) are reported in all the hydrodynamic simulations (Somov et al., 1981, Nagai and Emslie, 1984, Fisher et al., 1985 a-c) and widely investigated from the blue and red-shifted spectral measurements in UV and $H_{\alpha}$ emission, respectively. For some events, or some beam parameters, these can be nearly equal (e.g. see Zarro et al., 1988) while for many other beams if the electron Ohmic losses are taken into account (Zharkova and Gordovskyy, 2006) only the blue shifts could be observed without the noticeable red ones.

As can be seen from Figure \ref{hydr}, the hydrodynamic response to the injection of the pure electron or mixed proton/jet beams are substantially different. The electron beam injected during 10 s produces a smaller (by factor 2-2.5) temperature increase in the corona (top left plots), a smaller (by an order of magnitude) density depression of the coronal plasma into the chromosphere (the middle left plots) and smaller (by factor 2-2.5) evaporation velocities (the bottom left plots) compared to those by mixed proton/jet beams (the right plots in Figure \ref{hydr} and Figure \ref{hydr_shock}). 

In addition, the mixed proton/jet beam forms the lower temperature shock, which is spread much deeper into the lower chromosphere between the column depths of $2\times10^{20}\div 8\times10^{21}$ $cm^{-2}$ (see Figure \ref{hydr_shock}for a closer view of the shock in the first 100 s). The velocities of the shock induced by the mixed proton beam are also higher (by a factor of 2-2.5) than those induced by the pure electron beam.  These macro-velocities induced by the proton beam decrease in the region with a column density of $5\times10^{21}$ $cm^{-2}$ to a few km/s compared to those less than 1 km/s for the one induced by electrons.  The momentum induced by this shock is transferred to the photosphere within much shorter timescale under 1 minute because it is formed in much deeper and denser atmospheric levels. This also is confirmed by the temporal profile of the macro-velocity variations at the lower edge of the shock (the very right points in the distributions in Figure \ref{hydr_shock}) induced by proton beams shows an increase, to a few km/s, of the edge macrovelocities, within a minute, that resembles those measured in Figure \ref{vel_vert}. 

The proton induced shock deposits its momentum from the depths $5-8\times10^{21}$ $cm^{-2}$ to very dense plasma beneath that is delievered with a velocity of about 2 km/s through about 120 km of the solar atmosphere before approaching the column depth of $4\times10^{23}$ $cm^{-2}$ for Ni line region.  While the electron-induced shock is required to travel from the column depth of $(2-5)\times10^{20}$ $cm^{-2}$, or about 350 km, with the same or the twice lower velocity that will allow the momentum to reach the Ni region with a delay of 3-6 minutes.

\subsection{ The region for Ni 6768 line formation } \label{nikkel}
In order to understand the effect of the hydrodynamic shocks on the doppler measurements in Ni $6768 \AA$ line by the MDI instrument, one needs to establish the line formation region. This can be done by using the full non-LTE simulations (without magnetic field effects) of the Ni line in the ambient plasma with the coronal abundance of elements and molecules (up to 23 in total are considered, the main ones are CO, $C_2$, CH, CN)(Bruls, 1993; Uitenbroek, 2001; Zharkova and Kosovichev, 2002). The ambient plasma temperature, density and macrovelocities are described by the hydrodynamic solutions presented in \S~\ref{hd_mod}. 

We consider the two cases: the quiet Sun atmosphere outside sunspots (Figure \ref{ni_reg}, upper plot) before the beam onset and the flaring atmosphere heated by beam electrons (Figure \ref{ni_reg}, bottom plots). The contribution functions for the regions, where the Ni line $6768\AA$ is formed, are plotted by the grey curves with the grey arrows marking the maximum contribution. The background contribution functions (for the continuum) are plotted by the blue curves with the blue arrows marking a region of the maximum continuum contribution around the Ni line. The total contribution functions for all elements are plotted by the black curves. The red curves show the Plank contribution functions and the yellow lines present the mean intensity $J$ from a given atmospheric level for the temperature distributions taken from Figure \ref{hydr}(the top left plot, time = 7s after the beam injection). 

The formation region for the Ni line $6768\AA$ lies approximately within the column depths of $(2.0-4.0)\cdot10^{23}$ $cm^{-2}$, or around 200 km for the quiet Sun (Figure \ref{ni_reg}, upper plot) that increases to $(4.0-6.0)\cdot10^{23}$ $cm^{-2}$, or around 180 km, for the flaring atmosphere heated by beam electrons (Figure \ref{ni_reg}, bottom plot) that agrees rather well with the estimations found from the previous non-LTE simulations (Jones, 1989; Bruls, 1993; Zharkova and Kosovichev, 2002).

Let us compare these column depths with those obtained for the lower temperature shocks simulated from the hydrodynamic responses to injection of the electron or proton/electron beams as discussed in \S~\ref{hd_mod}. One can notice that the column depths for the Ni line region are closer to a hydrodynamic shock formed by the mixed proton/jet beam (Figure \ref{hydr_shock} and Figure \ref{hydr}, right plots). Also the shock velocities decrease to a few km/s towards the deeper depth edge (the right ends of each curve in Figure \ref{hydr_shock}) and this decrease has a temporal profile of the lower edge velocities increasing from 0.1 km/s (at 1s) to 2 km/s (at 50s) and then decreasing back similarly to those measured by MDI (see Figure \ref{vel_vert}). 

As discussed above in \S~\ref{hd_mod}, this shock needs only about 60 s to reach the Ni region that allows the detection of the seismic response nearly simultaneously with the hard X-rays as reported for this flare in \S~\ref{data} and \S~\ref{td_obs}. While the shock caused by the electron beam is formed much higher in a flaring atmosphere (Figure \ref{hydr}, left plots), it can be measured in the Ni $6768\AA$ line delayed by up to 6 min compared to the time of X-ray and $\gamma$-ray emission.
 
\subsection{The momenta delivered by shocks and the proposed scenarios} \label{comp}

In \S~\ref{hydro} we established that the hydrodynamic shock, or lower temperature condensation induced by the proton beams is denser and wider in depth moving twice as fast as those induced by the electron beam. The travel time to the photosphere of the shock caused by electron beams is about 180-360 s (compare the depth and densities of the lower-temperature condensations plotted in the lower graphs, in Figure \ref{hydr}) before it reaches the photosphere.  

The hard proton beam of a moderate power ($10^{12}$ $erg/cm^2/s$ mixed with the thermal jet protons  presented in Figure \ref{hydr} and Figure \ref{hydr_shock}) forms a shock much deeper in the lower chromosphere than the electron beam and it travels to the photosphere in about $60$ s. The shorter travel time of such a shock can explain the close correlation of the onset times for the seismic waves, measured in the TD diagrams (Figure \ref{sources}) with the hard X-rays and $\gamma$-ray emission bursts reported in \S~\ref{td_obs}. 

In order to confirm it, let us calculate the momenta delivered by the HD-shocks caused by either beam by using the method described in \S~\ref{hd_mom} and taking the particle densities, macro-velocities and duration times from Figure \ref{hydr} and the source areas from Table 2. 

The shock caused by a pure electron beam (Figure \ref{hydr}, left plots) with the beam parameters taken from Figure \ref{coronas} has a density about $5\times 10^{12}$ $cm^{-3}$, an average macro-velocity about $ 1.8\times 10^7$ cm/s and a duration about 100s. This shock is formed at the depth of about $(2-5)\times 10^{20}$ $cm^{-2}$. The momentum still requires a timescale $>180$s to be delivered to the Ni line formation region $(2-6)\times10^{23}$ $cm^{-2}$. Then for the source S1 with the area of about $5.05\times 10^{17}$ $cm^2$ the electron-formed shock can deliver a momentum of about $1.3\cdot 10^{19}$ $g \cdot cm /s$.  

Evidently, the shock produced by a pure electron beam does not contain enough momentum to account for the seismic responses recorded in either source. Also a time delay of about 180-360 s is required for this shock to reach the region of the Ni line formation to cause a delay in the seismic response appearance compared to the emission in hard X-rays and $\gamma$-rays. 

On the contrary, the parameters of the proton-formed shock are much more relevant to the explanation of the momenta observed in the seismic sources S1, S2 and S3. This shock occurs much closer to the region where the Ni line is formed, it contains much denser material (up to $10^{14}$ $cm^{-3}$) and higher macro-velocities (($2-3)\times10^7$ cm/s.  By substituting these parameters and the areas of the sources S1, S2 and S3 taken from Table 2 into the formula (18), one can find that this shock can deliver the much higher($>4.4\times10^{22}$  $g cm/s$) momentum compared to that delivered by the electron-formed one. The momentum induced by this shock is transferred to the photosphere within much shorter timescale under 1 minute because it is formed in much deeper and denser atmospheric levels. Therefore, the magnitudes of the momentum carried by the shock caused by the mixed proton/jet beam and the timescale, within which it reaches the region of Ni line, are rather close to those deduced from the TD-diagrams  discussed in \S~\ref{est_mom} (Table 2) for all the seismic sources.  

Since $\gamma$-ray emission by high energy protons was observed by KORONAS only after 11:06:00 UT (Kuznetsov et al., 2006), when the sources S2 and S3 appeared, one can assume that these sources are produced by a high energy proton beam combined with the lower energy ($<200 MeV$) quasi-thermal protons of the separatrix jets. They together  can deliver to the photosphere the momenta via the hydrodynamic shock, to cause directly the observed seismic responses and to energise ambient electrons to energies high enough for the hard X-ray and $\gamma$-ray emission observed at 11:06:00 UT in the third burst as discussed in \S~\ref{hxr}. 

For the source S1 sufficient momentum can be delivered by a very high energy electron beam, which is reported to have energies of hundreds of MeVs (Kuznetsov et al., 2006) and occurred at 11:02:00 UT in the first burst (Figure 3) combined with the lower energy ($<200 MeV$) quasi-thermal jet protons.  Then the hydrodynamic response to such heating could lead to a shock formed slightly deeper in the flaring atmosphere than those for the electron beam from Figure \ref{hydr} and slightly higher than for the protons. Within 3-4 minutes after the beam onset the momentum caused by this hydrodynamic shock can reach the Ni region and cause the observed seismic waves as discussed above which explains the delay of about 3-4 minutes between the seismic response in the source 1 and the hard X-ray emission in the first burst started at 11:02:00 UT.

\section{Conclusions} \label{cons}
	
In the current paper we report the 11 sources with downward motions higher than 1 km/s in the flare 28 October 2003 and the seismic waves in 3 of them (S1-S3) detected with the time-distance (TD) diagram technique. The 3 seismic sources started around 11:05:00-11:06:00 UT, had slightly different downward (vertical) velocities and heights of ridges, or horizontal velocities, pointing out to different agents causing them. 

We investigate a few agents able to deliver the required energy and momentum to the flaring atmosphere: power law electron beam, power law proton beam and quasi-thermal protons with energies below $200$ MeV occurring through acceleration by a super-Dreicer electric field in reconnecting current sheets on the top of a flaring atmosphere (Zharkova and Gordovskyy, 2004, 2005a). Electron beams are considered to deposit their energy in collisions and Ohmic dissipation, proton  beams - in collisions and generation of kinetic Alfven waves with their following dissipation in Cherenkov's resonance (Gordovskyy et al., 2005). 

The hydrodynamic responses  to a precipitating power law electron beam or to a proton beam mixed with the protons of separatrix jets are also investigated. The hydrodynamic response to the injection of the pure electron or mixed proton/jet beams are found to be substantially different. The electron beam injected during 10 s produces a smaller (by factor 2-2.5) temperature increase in the corona (top left plots), a smaller (by an order of magnitude) density depression of the coronal plasma into the chromosphere (the middle left plots) and smaller (by factor 2-2.5) evaporation velocities (the bottom left plots) compared to those inducedcby mixed proton/jet beams (the right plots in Figure \ref{hydr} and Figure \ref{hydr_shock}). 

\begin{figure}  
\includegraphics[scale=0.45]{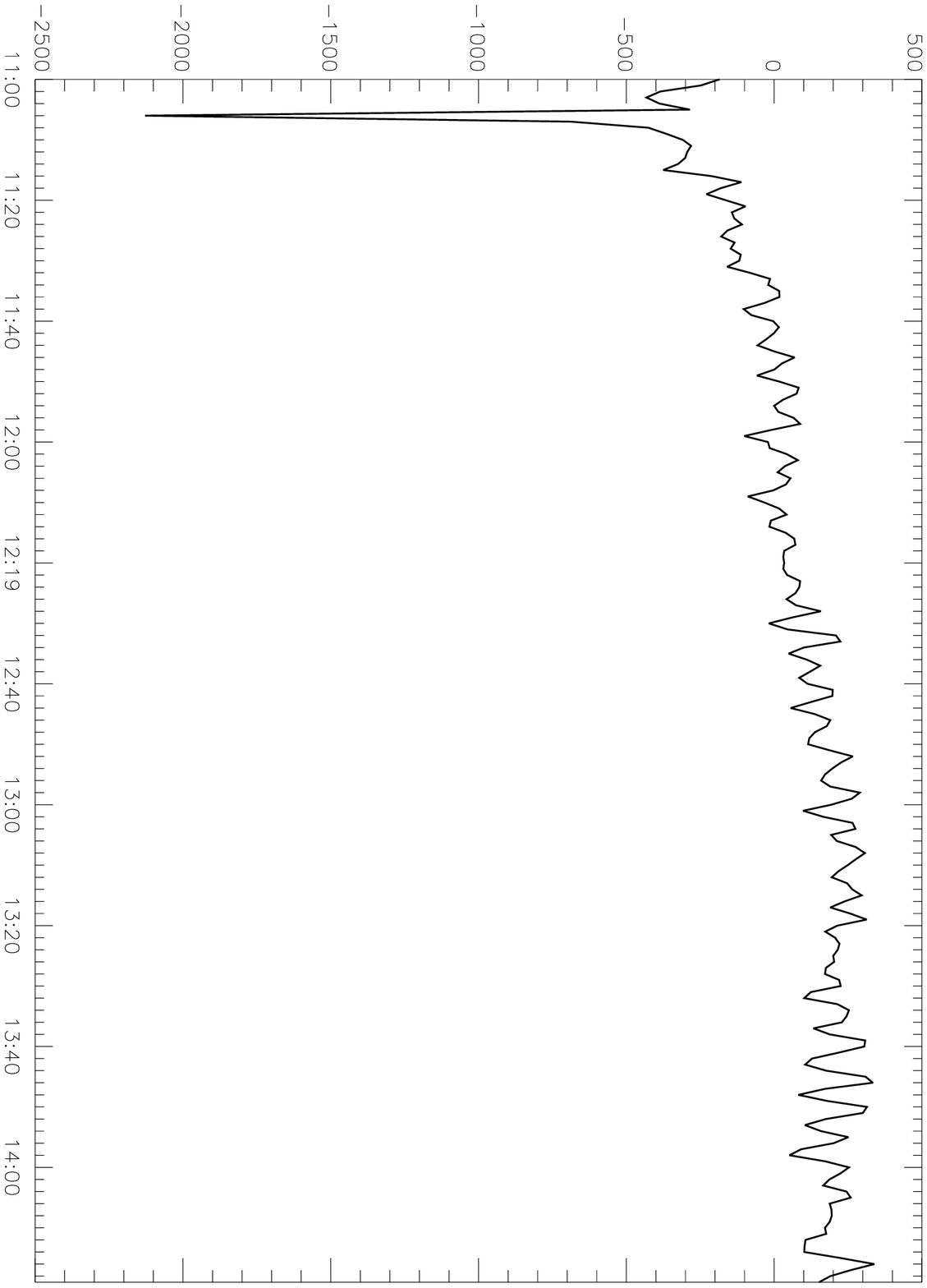} 
\includegraphics[scale=0.45]{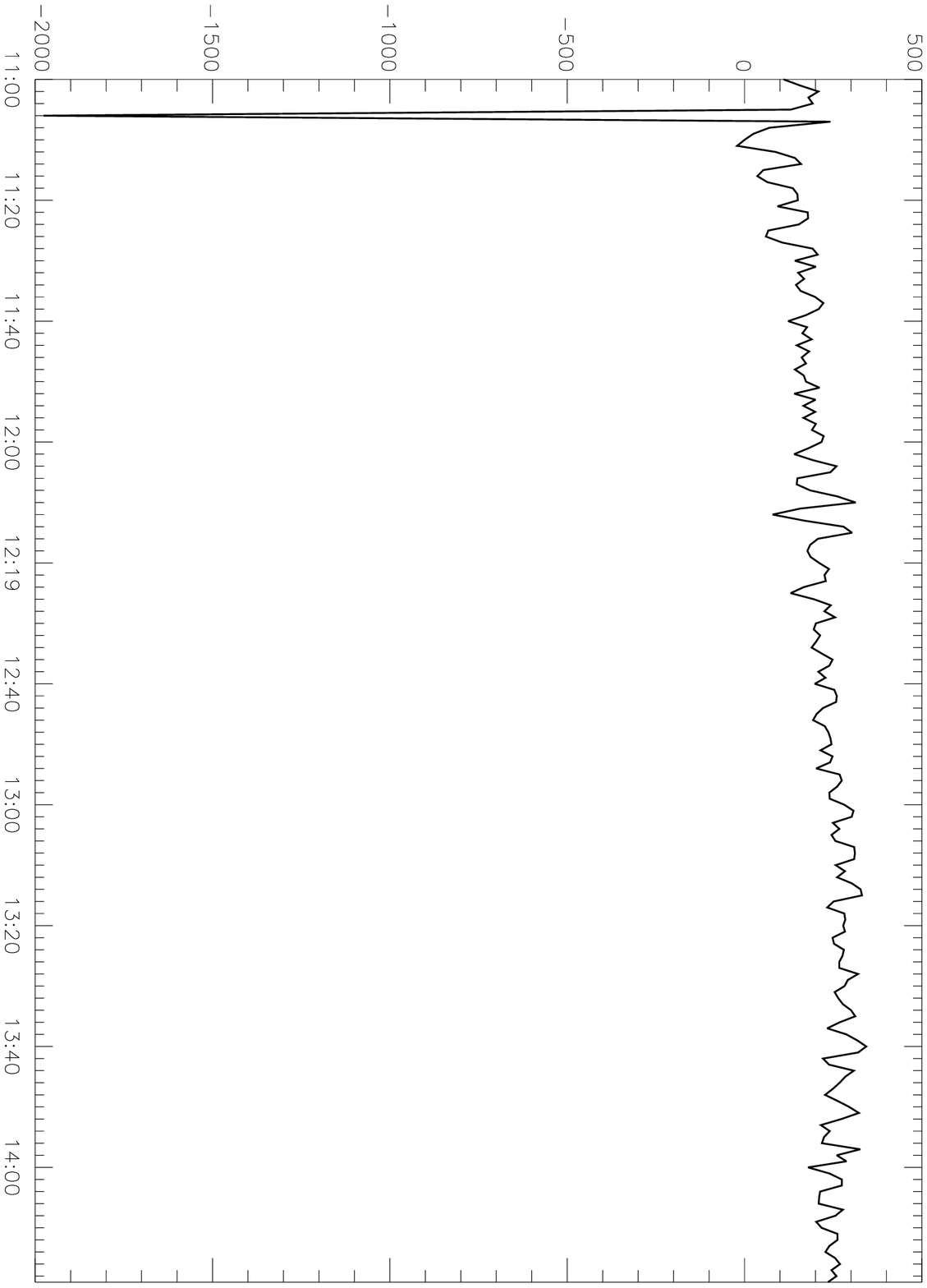}  
\includegraphics[scale=0.45]{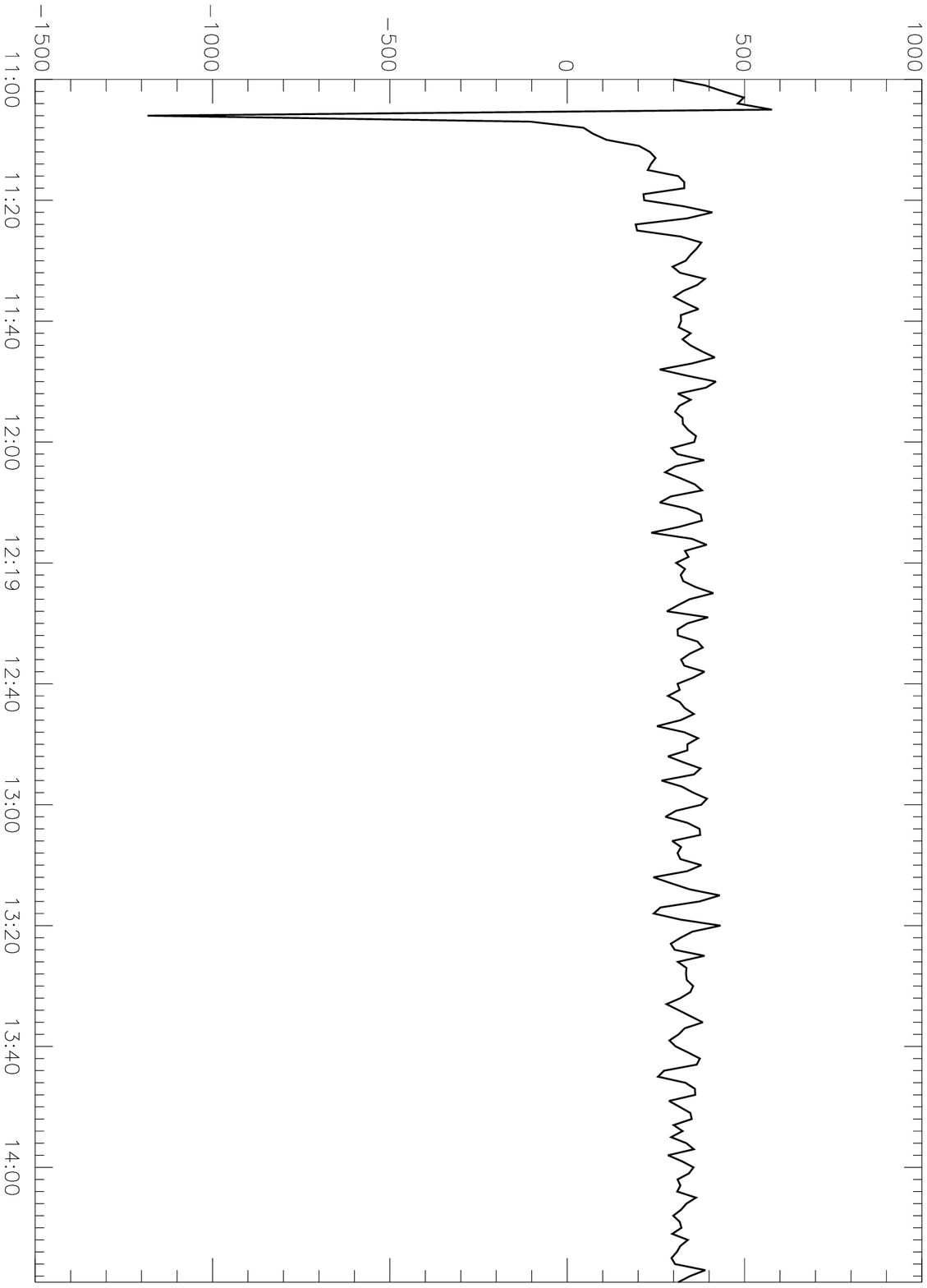} 
\caption{ The downward velocities (m/s) (Y-axis) versus time (minutes) (X-axis) in S1 source (upper right plot), S2 (upper left plot) and S3 (bottom left plot) in the centre of gravity locations with the downward Doppler motions depicted in Figure \ref{dop11}b and Table 1.} 
\label{vel_vert} 
\end{figure}

\begin{figure}
\begin{center}
\includegraphics[scale=0.55]{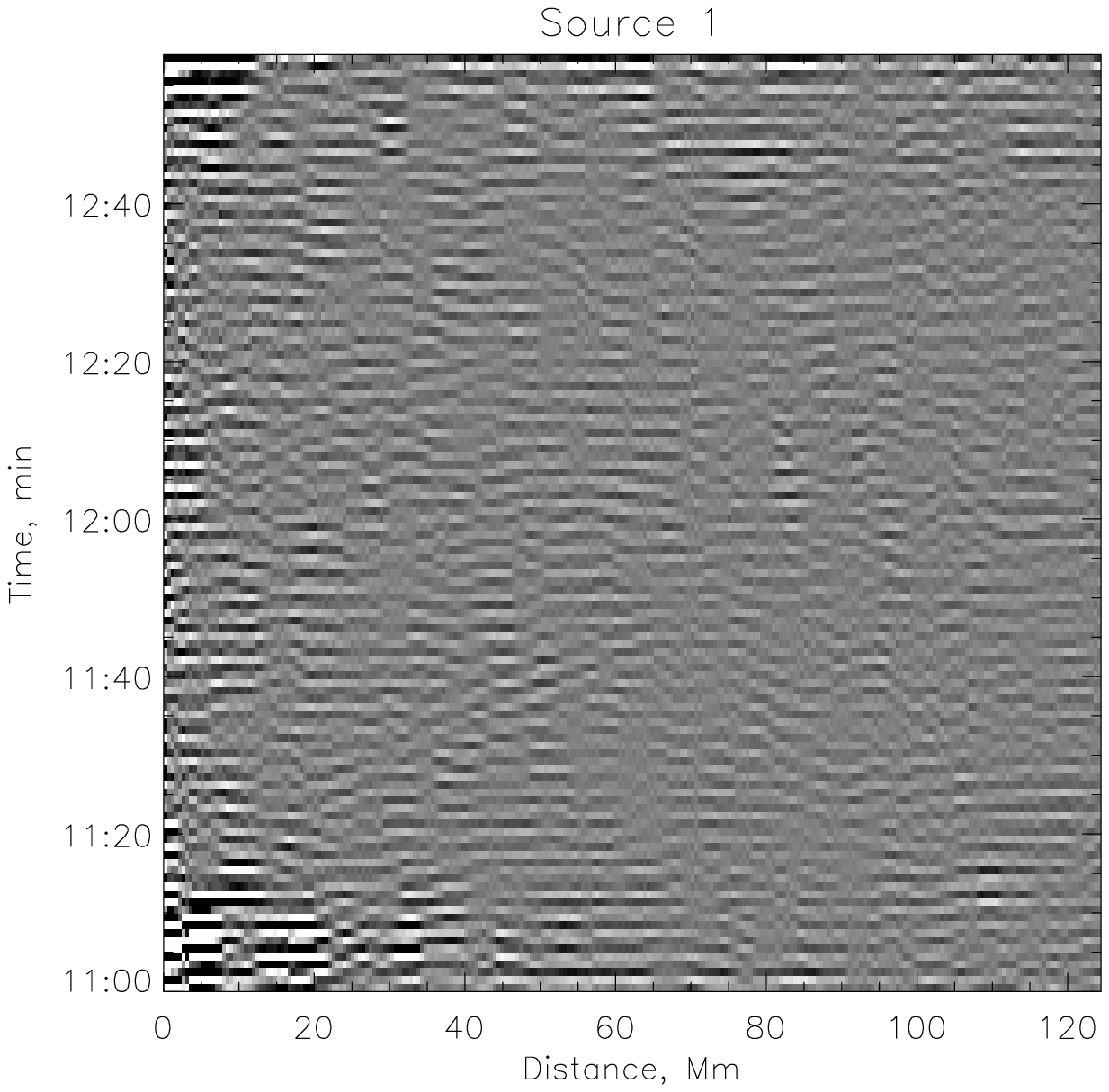}\includegraphics[scale=0.55]{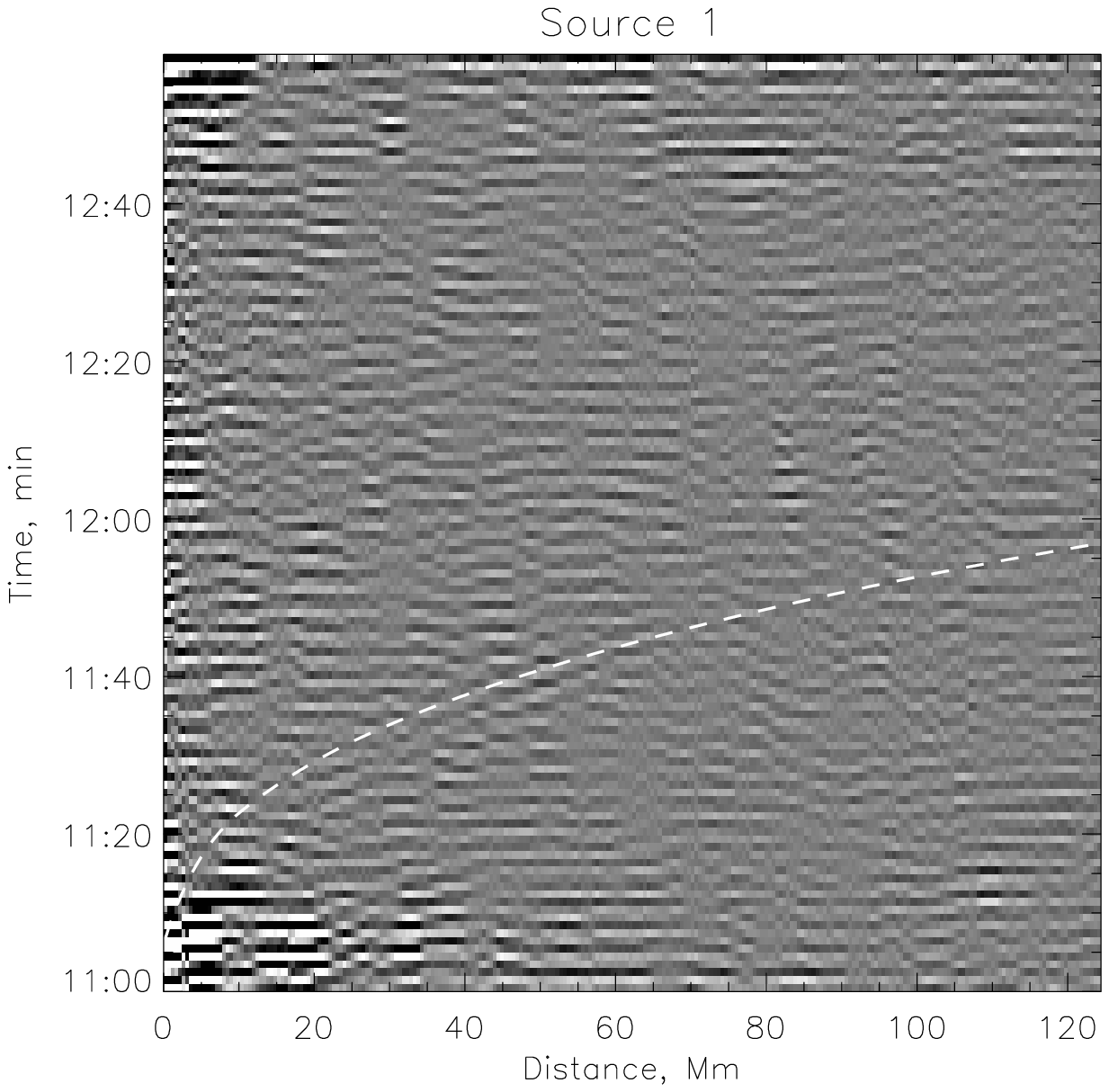}
\includegraphics[scale=0.55]{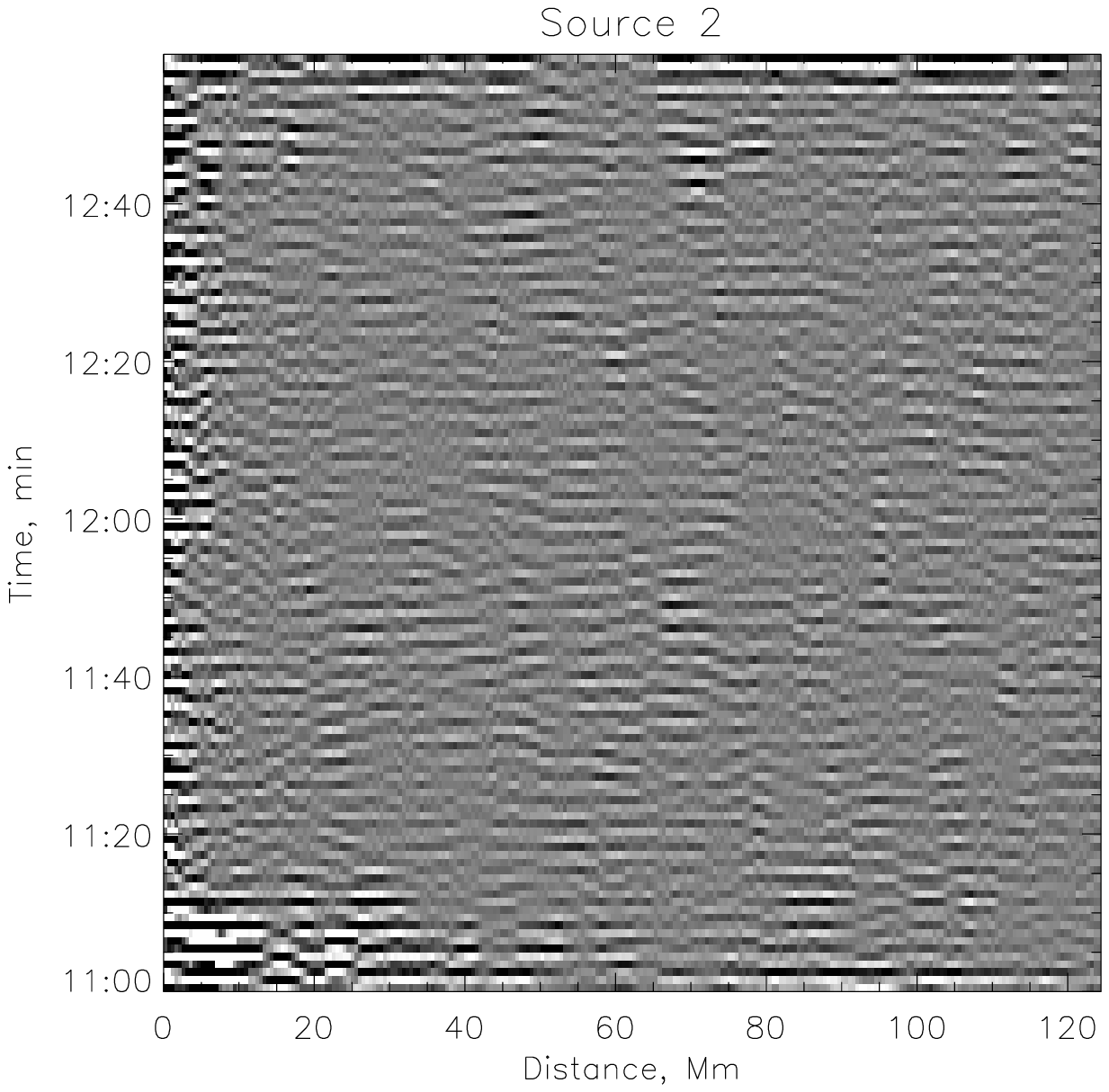}\includegraphics[scale=0.55]{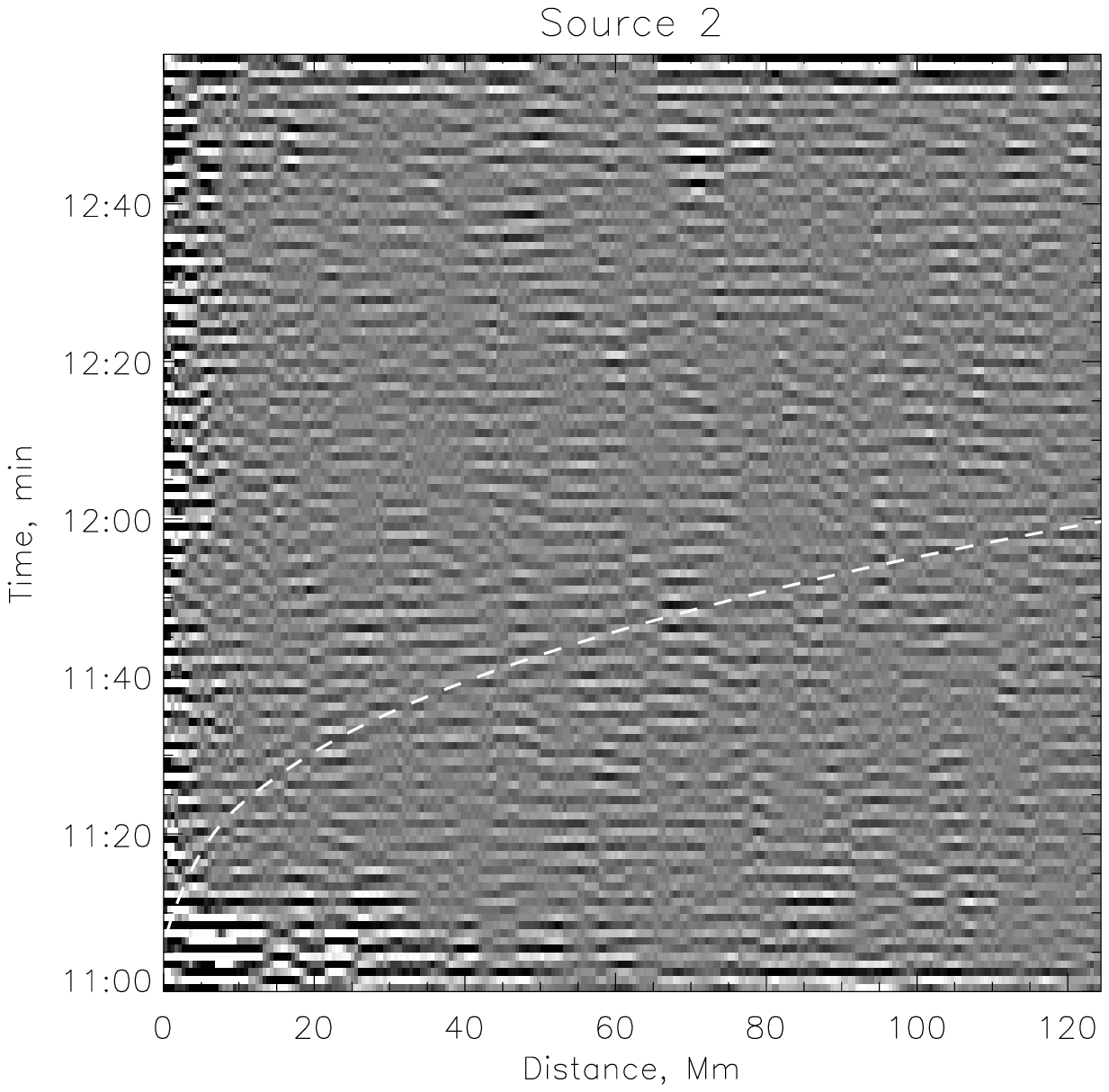}
\includegraphics[scale=0.55]{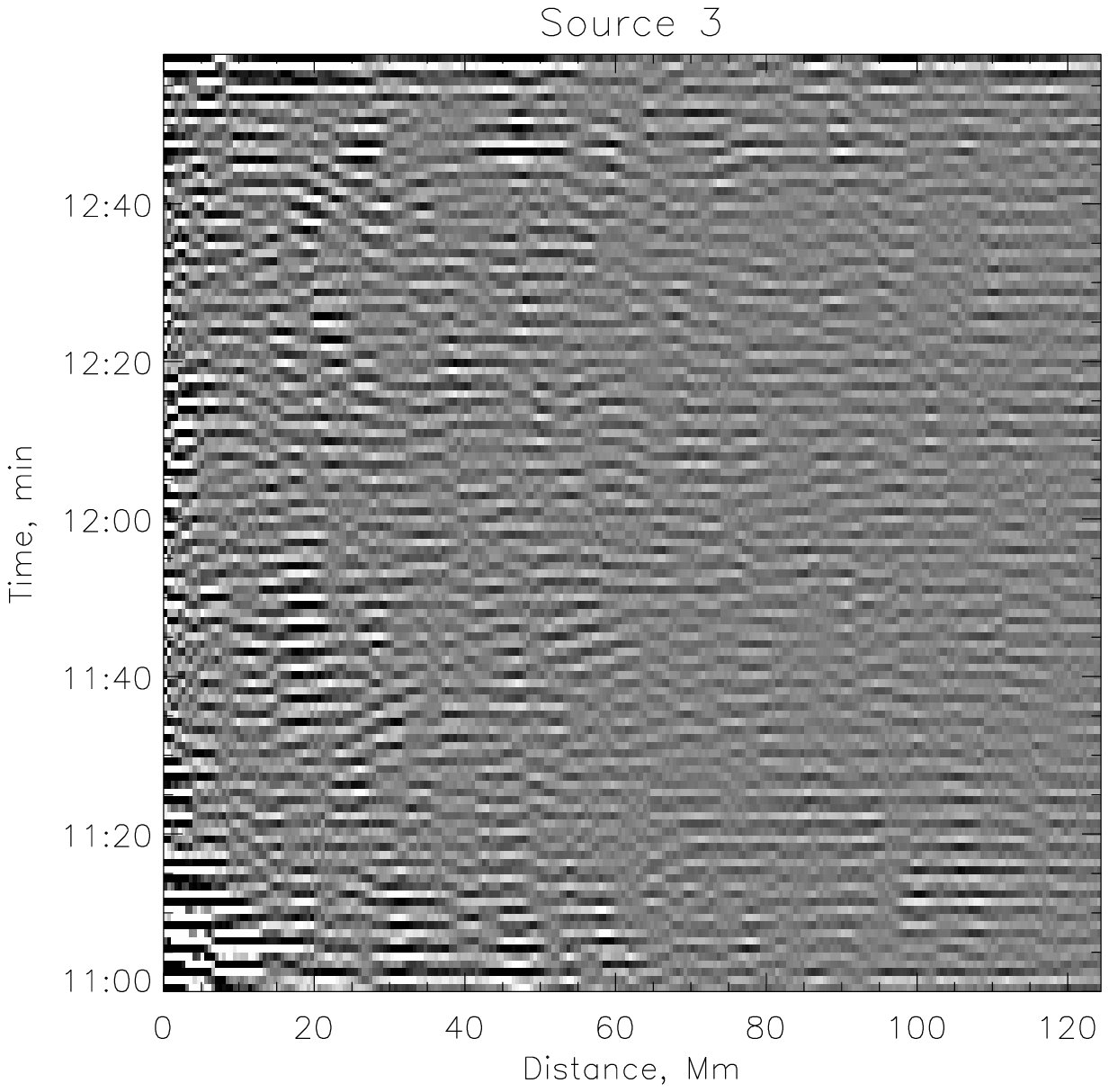}\includegraphics[scale=0.55]{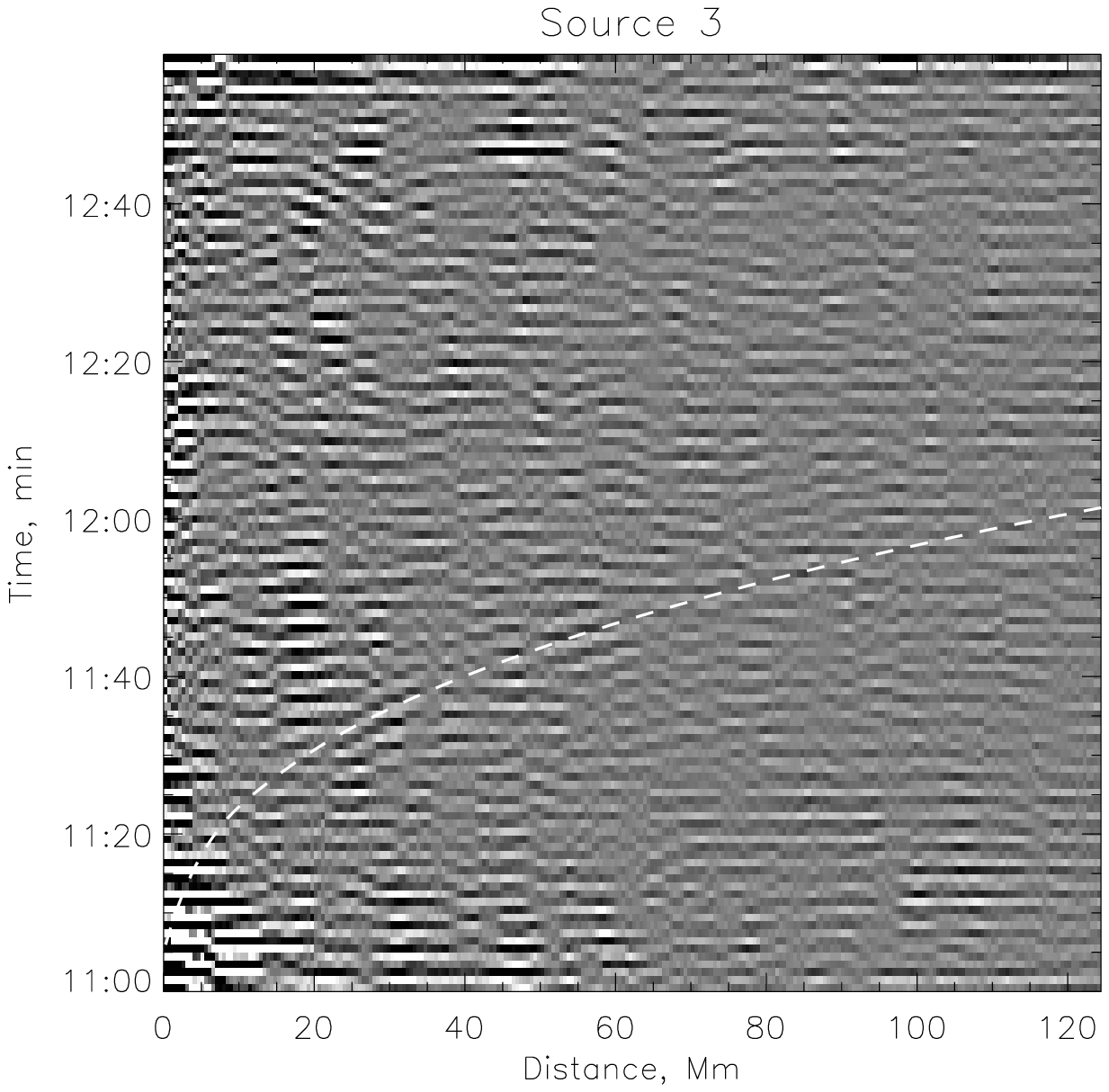}
\end{center}
\caption{The frequency-filtered time-distance diagrams for the seismic sources: S1 without a ray path (left upper plot) and S1 with the ray path (right upper plot ), similar for the sources S2 (middle plots) and S3 (bottom plots).}
\label{sources}
\end{figure}

\begin{figure} 
\includegraphics[scale=0.8]{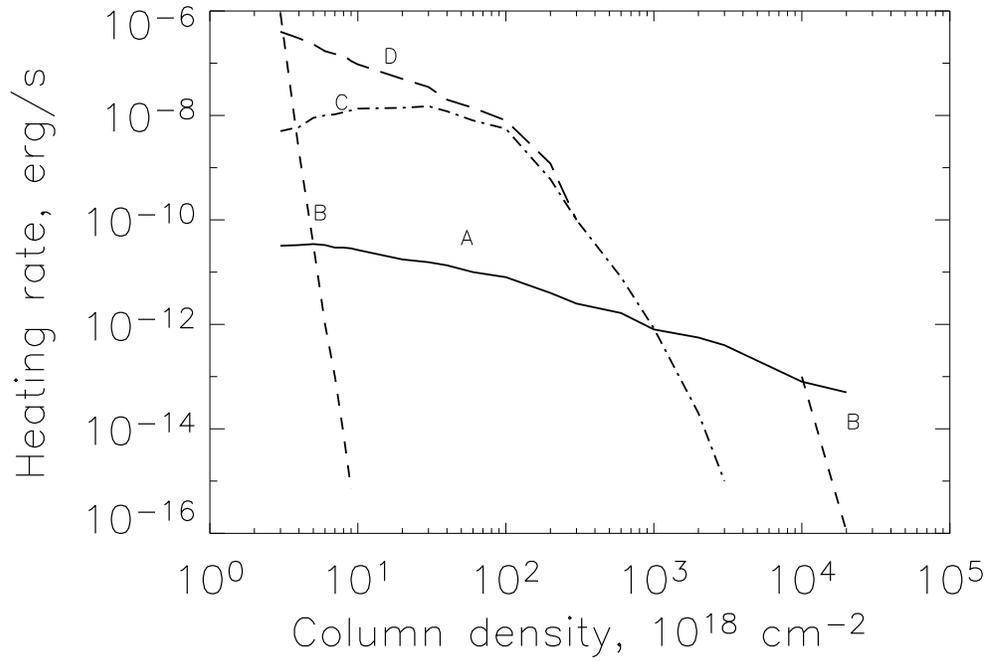} 
\caption{The heating rates for different populations of accelerated particles: the curve 'A' corresponds to a weak and hard electron beam, the curve "B" marks the heating by "slow" jet protons; "C" to a strong and soft electron beam and "D" to a "fast" proton beam.}
\label{heat_g} 
\end{figure}

\begin{figure}
\includegraphics[scale=1.35]{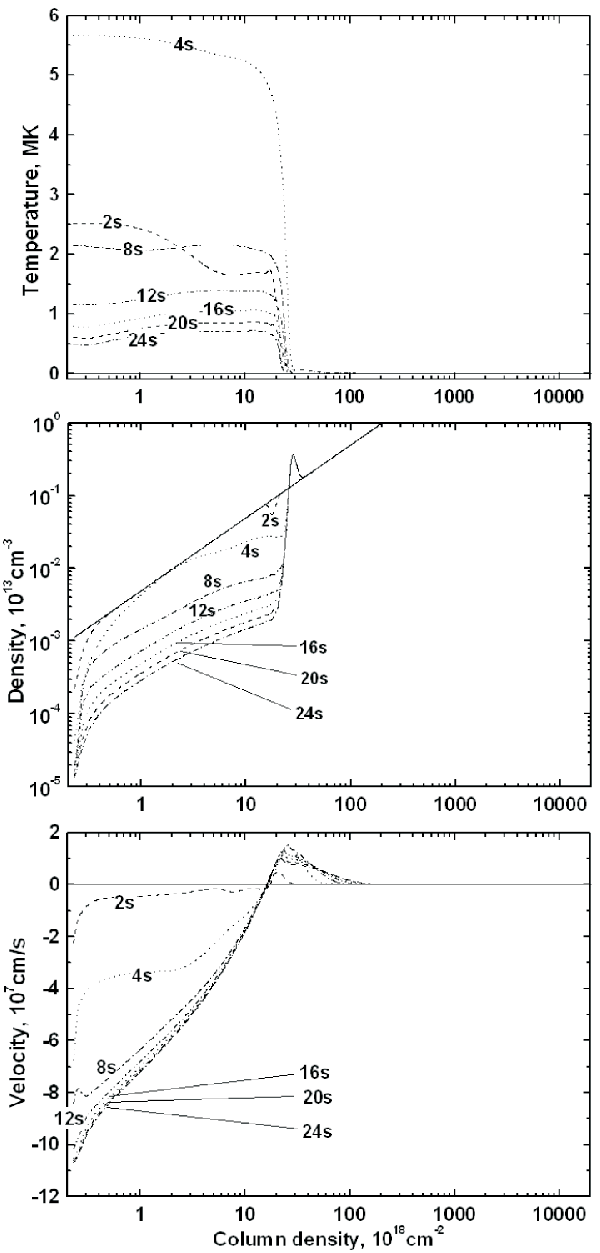}
\includegraphics[scale=1.35]{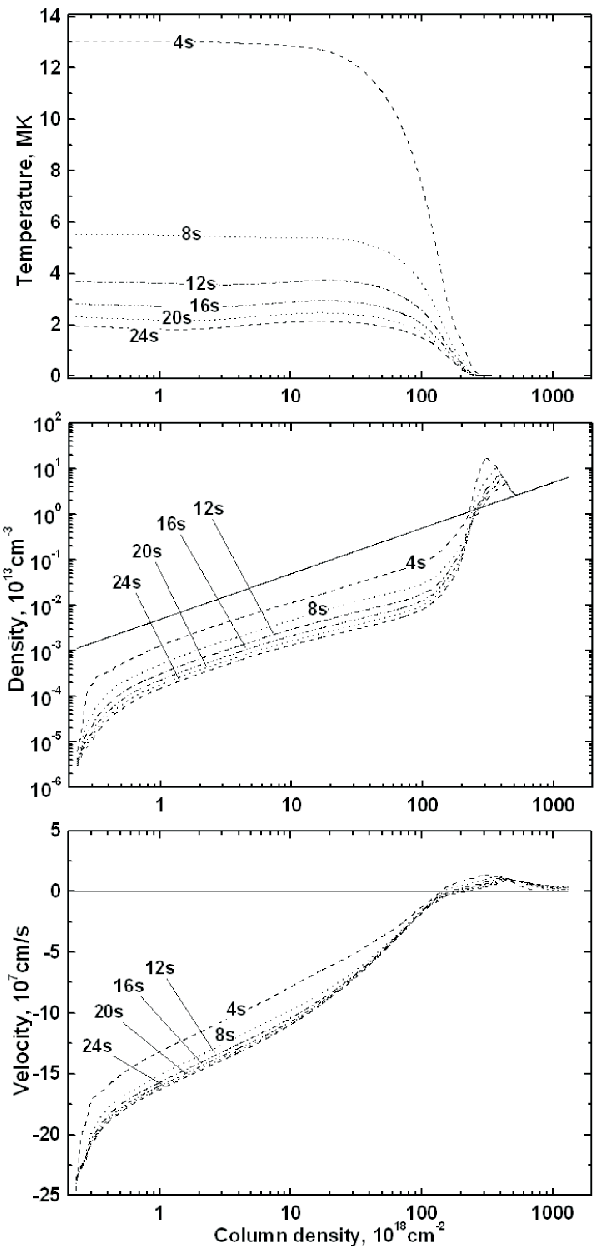}
\caption{The hydrodynamic responses of a flaring atmosphere caused by pure electrons (the left column) or by mixed proton and electron beams (the right column) with the parameters of protons deduced from $\gamma$-rays and electrons from X-rays from RHESSI and KORONAS. The upper plots present the ambient electron temperatures, the middle ones - densities and the lower plots - macro-velocities, the numbers on the graphs show the times in seconds after the beam injection. }
\label{hydr}
\end{figure}

\begin{figure}
\includegraphics[scale=1.0]{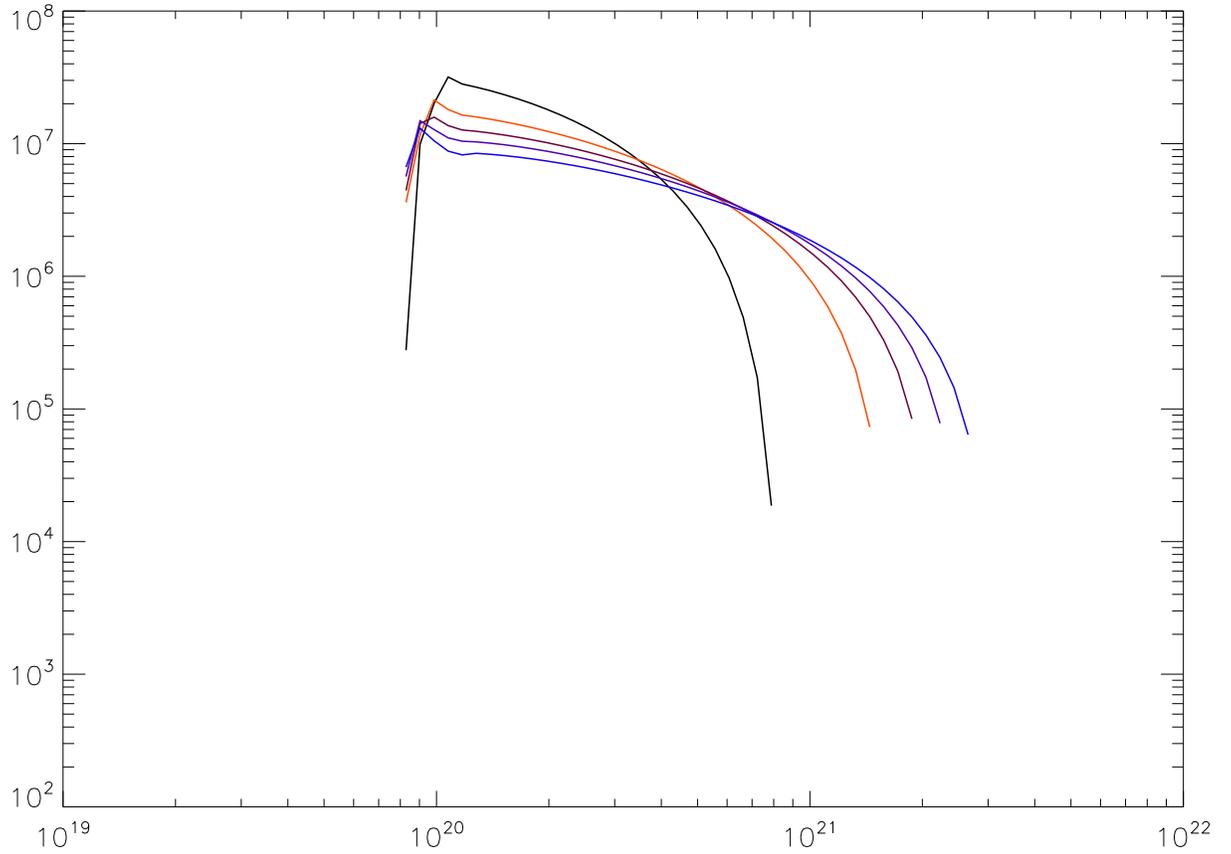}

\caption{The close-up from Figure \ref{hydr} of the temporal variations of macro-velocity in $cm/s$ in a lower temperature condensation, or a hydrodynamic shock (Y-axis) versus the column depth in $cm^{-2}$ (X-axis), appeared in response to the injection of mixed proton/electron beams (Figure \ref{hydr}, right plots(1- after 10s (black line), 2 - 30s (grey line), 3 - 50s (purple line), 4 - 70s (blue line) and 5 - 100s (green line). }
\label{hydr_shock}
\end{figure}

\begin{figure} 
\includegraphics[scale=0.75]{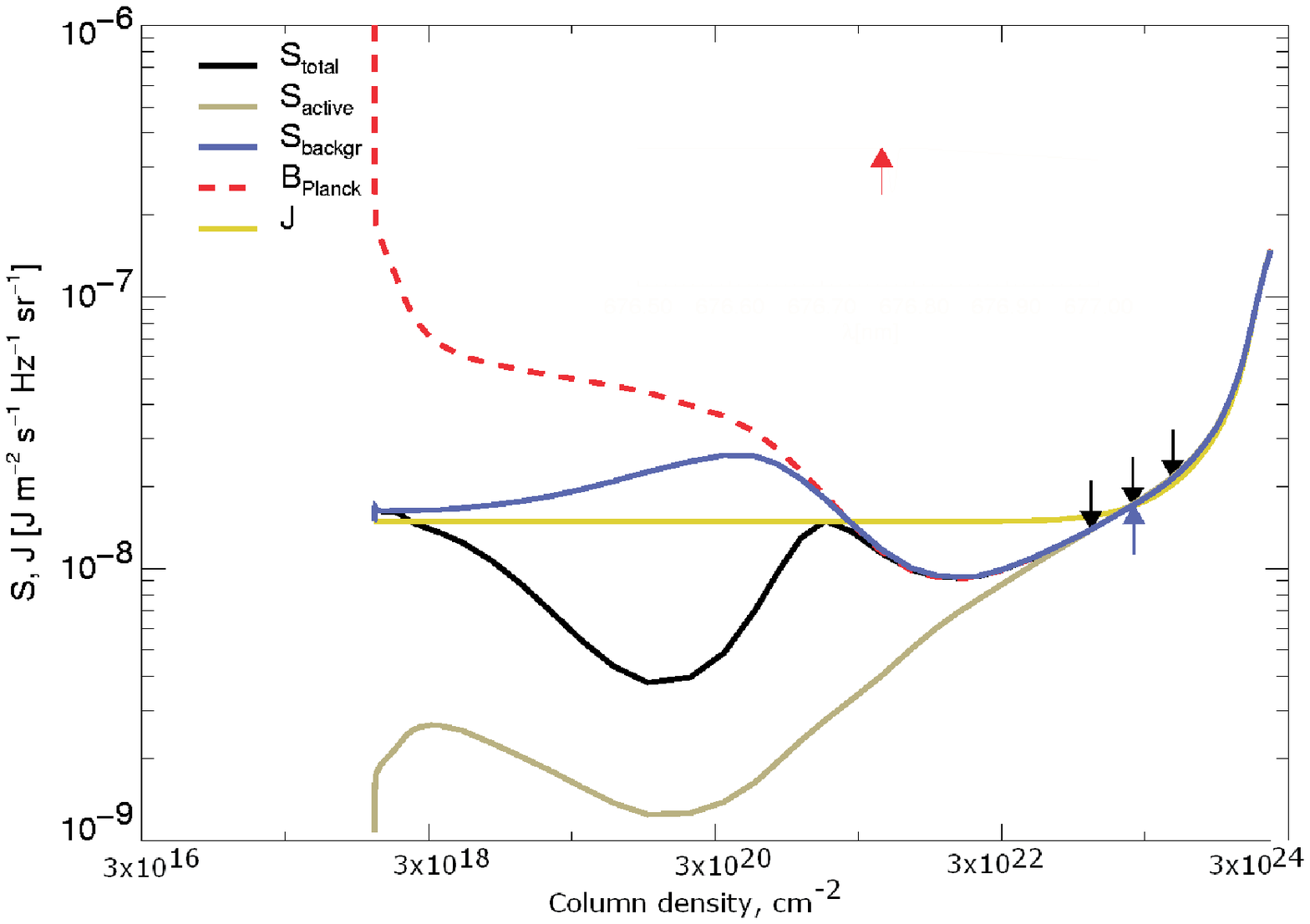} \includegraphics[scale=0.75]{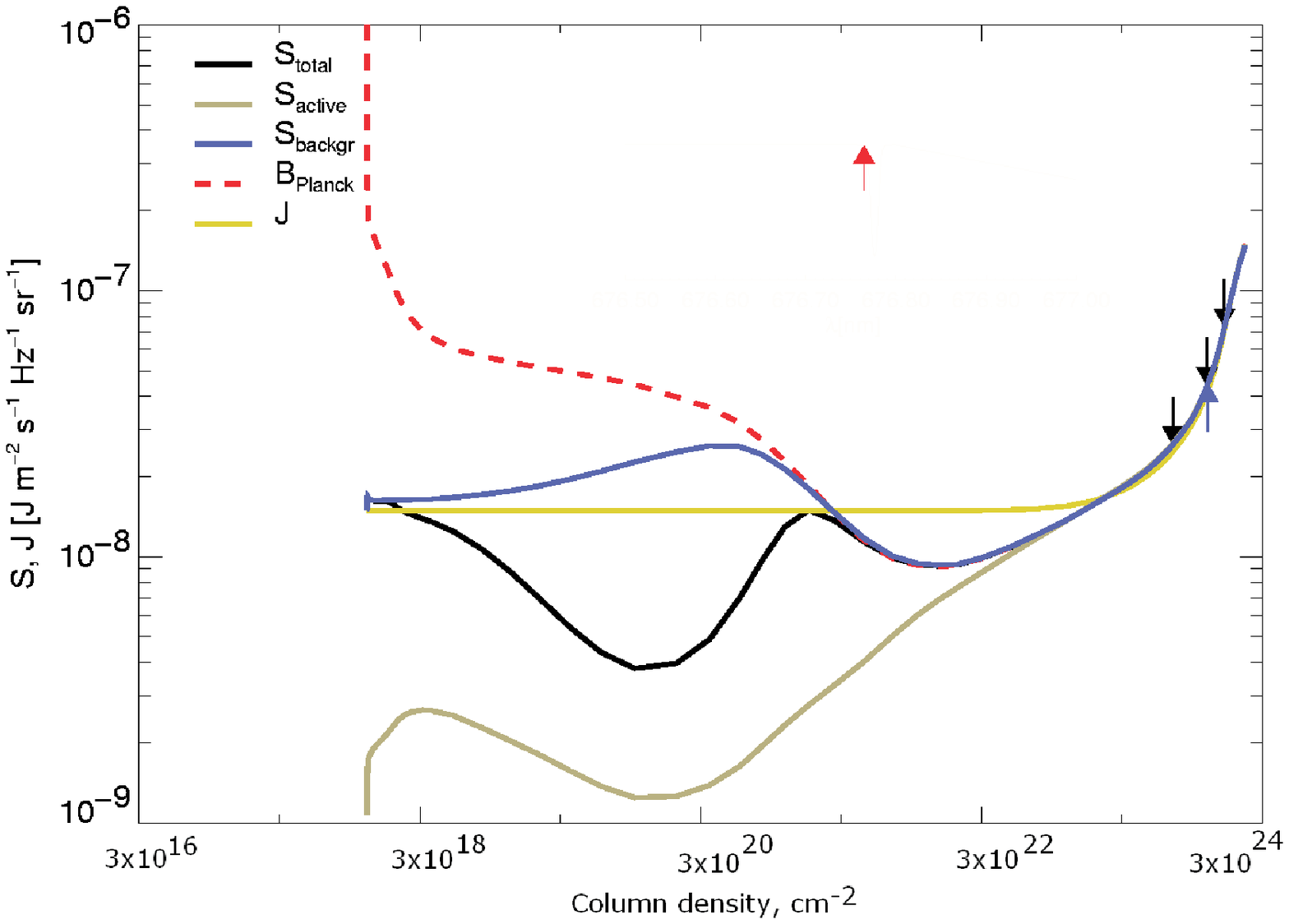}
\caption{The source function distributions calculated for the quiet atmosphere (top plots) and the flaring atmosphere (bottom plots) heated by a hard intense electron beam ($\gamma$=3, $F_0=10^{12}$ $erg/cm^2/s$ for the Ni line transition 6768A ($S_{active}$, grey line), background elements ($S_{backgr}$, blue line), Plank function ($S_{plank}$, red dashed line) and total for all elements ($S_{total}$, black line) as well as the mean intensity $J$ (yellow line) simulated using the full NLTE MULTI-based) approach for the full coronal abundance of elements and some molecules (CO, $C_2$, CH, CN, 23 in total)  (Uitenbroek, 2001; Zharkova and Kosovichev, 2002).}
\label{ni_reg}
\end{figure}

In addition, the mixed proton/jet beam forms a low temperature shock, which is spread much deeper into the lower chromosphere between the column depths of $2\times10^{20}\div 8\times10^{21}$ $cm^{-2}$ (see Figure \ref{hydr_shock} for a closer view of the shock in the first 100 s). The velocities of the shock induced by a mixed proton beam are also higher (by a factor of 2-2.5) than those induced by a pure electron beam.  These macro-velocities induced by the proton beam decrease in the region of a column density of $5\times10^{21}$ $cm^{-2}$ to a few km/s compared to those less than 1 km/s for the one induced by electrons. Also the temporal profile of the macro-velocity variations at the lower edge of the shock (the very right points in the distributions in Figure \ref{hydr_shock}) induced by proton beams shows an increase to a few km/s of the edge macrovelocities within a minute that resembles those measured in Figure \ref{vel_vert}. 

The momenta and start times measured from the TD diagrams in the sources S1-S3 are compared with those delivered to the photosphere by different kinds of high energy particles with the parameters deduced from hard X-ray and $\gamma$-ray emission as well as by the hydrodynamic shocks caused by these particles.

The energetic protons (power laws combined with Maxwellian ones from the separatrix jets) are shown to deliver momenta high enough and to form hydrodynamic shocks much deeper in a flaring atmosphere compared to a pure electron beam. This allows the proton-formed shocks to travel to the photosphere shorter distances and less time resulting in seismic waves occurring nearly simultaneously with the high energy emission as observed in the sources S2 and S3. The source S1 is likely to be associated with a hard power law electron beam mixed with the quasi-thermal protons of the separatrix jets.

\acknowledgements
 The authors thank Drs. G. Share of the University of Maryland, R.Murphy of Naval Research Laboratory, US and Dr. V.Kurt of Moscow University, Russia for kindly offerred hard X-ray and $\gamma$-ray spectra from RHESSI, INTEGRAL and KORONAS and Dr. C.Lindsey of NW Research Associates, Boulder for a general discussion and our anonymous referee for his constructive comments from which the paper strongly benefited. SZ also acknowledges that this work is being partially supported by the UK Particle Physics and Astronomy Research Council.

\clearpage

\begin{table}   
\begin{center}   
\caption{ The heliographic locations $L_0$ and $B_0$ of the 3 seismic sources detected by the TD technique (columns 1-3) and 4 seismic sources in the flare 28 October 2003 reported from the  holographic method \cite{doli05} (columns 4-6).}
\vspace{1em}  
    \renewcommand{\arraystretch}{1.2}  
    \begin{tabular}[h]{lrclcc}  
      \hline 
      Our sources  & $L_0$ $^\circ$ & $B_0$ $^\circ$ & DL05 &  $L_0$ $^\circ$  & $B_0$ $^\circ$  \\ 

      \hline  
      {$S \,1$} & 287.28 &  -15.96 & {$X \,3$} & 287.05 &  -15.78 \\  
      {$S \,2$} & 284.72 &  -17.62 & {$X \,1$} & 285.45 &  -17.61 \\  
      {$S \,3$} & 291.00 &  -16.64 & {$X \,4$} & 291.46 &  -16.43 \\  
             &  &  &  {$X \,2$} & 285.86 &  -16.01 \\ 

      \hline \\   
      \end{tabular}   
    \label{tab:table_01}   
\end{center}   
\end{table}  

\clearpage

\begin{table}  
\begin{center}   
\caption{ The areas $A$, downwards (vertical) velocities $V_{vert}$,  the duration of an observed downflow motions in the source location, $T_{impulse}$, the average  horizontal velocity  $V_{horiz}$ and the $Momenta$ deduced for the flare 28  October 2003 from the time-distance diagram for the 3 seismic sources S1, S2, S3 (Figure \ref{sources}). } 

\vspace{1em}  
    \renewcommand{\arraystretch}{1.2}  
    \begin{tabular}[h]{lrclcc}  
      \hline  
      Our sources  & $area, A$, $cm^2$ & $V_{vert}$, $cm/s$ , & $T_{impulse}$, s & $V_{horiz}$, $cm/s$ &  $Momenta, g\cdot cm/s$   \\  
      \hline  
      {$S \,1$} & $5.05\times 10^{17}$ &  $2.15\times 10^5$ & 90 & $40.8\times 10^5$ & $4.1\times10^{22}$  \\  
      {$S \,2$} & $3.34\times 10^{17}$ &  $2.00\times 10^5$ & 70 & $38.0\times 10^5$ & $3.8\times10^{22}$   \\  
      {$S \,3$} & $2.22\times 10^{17}$ &  $1.75\times 10^5$ & 70 & $35.4\times 10^5$ & $3.1\times10^{22}$   \\  
            
      \hline \\   
      \end{tabular}   
    \label{tab:table_02}   
\end{center}   
\end{table}   

\clearpage

\clearpage

\clearpage

\clearpage

\clearpage

\clearpage

\clearpage

\clearpage

\clearpage

\clearpage 

\end{document}